\documentclass[showpacs,aps,prd,reprint,superscriptaddress,nofootinbib,longbibliography]{revtex4-1}
\usepackage[colorlinks=true, pdfstartview=FitV, linkcolor=magenta,citecolor=blue, urlcolor=magenta,
bookmarks=true, bookmarksnumbered=true]{hyperref}
\usepackage[dvipdfmx]{graphicx}
\usepackage{amsmath,amssymb,bm,color,longtable,mathrsfs,amsfonts,slashed,ulem}

\newcommand \beq{\begin{eqnarray}}
\newcommand \eeq{\end{eqnarray}}

\newcommand{\Nc}{N_{\rm c}}
\newcommand{\Nf}{N_{\rm f}}

\newcommand{\lqcd}{\Lambda_{\rm QCD}}
\newcommand{\vp}{ {\bm p}}
\newcommand{\vq}{ {\bf q}}
\newcommand{\vecr}{ {\bm r} }
\newcommand{\vL}{ {\bm L} }

\newcommand{\vk}{ {\bm k}}
\newcommand{\vl}{ {\bm l}}

\newcommand{\vP}{{\bm P}}
\newcommand{\vx}{ \bm{ {x}} }

\newcommand{\vR}{ {\bm R}  }

\newcommand{\la}{\langle}
\newcommand{\ra}{\rangle}

\newcommand{\calL}{\mathcal{L}}

\newcommand{\calI}{\mathcal{I}}

\newcommand{\calN}{\mathcal{N}}

\newcommand{\calK}{\mathcal{K}}
\newcommand{\calJ}{\mathcal{J}}
\newcommand{\calV}{\mathcal{V}}
\newcommand{\calP}{\mathcal{P}}

\newcommand{\rmd}{\mathrm{d}}
\newcommand{\rmi}{\mathrm{i}}
\newcommand{\rme}{\mathrm{e}}

\newcommand{\up}{\uparrow}
\newcommand{\down}{\downarrow}

\begin{document}
\begin{flushright}
\end{flushright}

\title{Meson resonance gas in a relativistic quark model: \\
scalar vs vector confinement and semishort range correlations}

\title{Relativistic quarks in a meson resonance gas: \\
scalar vs vector confinement and semishort range correlations}

\author{Toru Kojo}
\email{toru.kojo.b1@tohoku.ac.jp}
\affiliation{Department of Physics, Tohoku University, Sendai 980-8578, Japan}

\author{Daiki Suenaga}
\email{daiki.suenaga@riken.jp}
\affiliation{Strangeness Nuclear Physics Laboratory, RIKEN Nishina Center, Wako 351-0198, Japan}
\affiliation{Research Center for Nuclear Physics, Osaka University, Ibaraki, 567-0048, Japan}

\date{\today}

\begin{abstract}
Smooth transitions from hadronic matter to hot and dense matter of quantum chromodynamics accompany continuous transformations in effective degrees of freedom.
The microscopic descriptions should include relativistic quarks interacting inside of hadrons.
In this work we construct a schematic constituent quark model with relativistic kinematics which captures the global trends of meson spectra in the light, strange, and charm quark sectors.
We examine the roles of the scalar- and vector-confining potentials as well as semishort range correlations in estimating the strength of central, spin-spin,  and spin-orbit interactions.
The quark dynamics in low-lying mesons is very sensitive to relativistic kinematics and short range interactions, 
while in high-lying mesons are sensitive to the composition of scalar- and vector-confinement.
After expressing mesons in terms of quark wave functions, we use them to describe the quark occupation probability in a meson resonance gas, and discuss how it can be related to its counterpart  in a quark-gluon-plasma.

\end{abstract}


\maketitle

\section{Introduction}
\label{sec:Introduction}

Quarks and gluons in quantum chromodynamics (QCD) play multiple roles in hadron physics;
they are not only constituents of hadrons, 
but also are mediators of hadron-hadron interactions \cite{tHooft:1973alw,Witten:1979kh,Fukushima:2020cmk}.
The quark exchanges are microscopic descriptions of traditional meson exchanges \cite{Yukawa:1935xg,PhysRev.106.1366,PhysRev.119.1784,PhysRevC.49.2950,PhysRevC.53.R1483},
and also can describe the baryon-baryon hard core repulsions at short distance \cite{Hatsuda:2018nes,Park:2019bsz,Oka:1980ax,Oka:1981rj,Oka:1981ri}.
When many hadrons are strongly interacting in a matter, hadrons are supposed to exchange substantial amount of quarks and gluons, 
and therefore it should be difficult to differentiate such strongly interacting hadronic matter from matters of quarks and gluons \cite{McLerran:2007qj,Masuda:2012kf,Kojo:2014rca}.

There are several circumstances where hadrons in a matter interact strongly.
In a heated hadronic matter many hadron resonances are generated and begin to overlap around $T\sim 150$ MeV,
and the matter smoothly transforms into a quark-gluon-plasma (QGP) through a crossover transition.
Such crossover has been confirmed by lattice QCD simulations \cite{Aoki:2006we} and fluctuation analyses in heavy-ion collisions \cite{STAR:2021rls}.
Another example of strongly interacting hadronic matter 
is a highly compressed nuclear matter in which nuclear many-body forces rapidly become important around twice nuclear saturation density \cite{Akmal:1998cf,Gandolfi:2011xu}.
The quark exchange picture of nuclear forces motivates the scenario of smooth transitions from nuclear to quark matter 
\cite{McLerran:2007qj,Masuda:2012kf,Kojo:2014rca,Fukushima:2020cmk}, 
which results in new qualitative features for equations of state such as the sound velocity peak in the crossover domain \cite{Masuda:2012kf,Masuda:2012ed,Kojo:2014rca,Baym:2017whm,Baym:2019iky,Kojo:2021wax,McLerran:2018hbz,Jeong:2019lhv,Duarte:2020kvi,Duarte:2020xsp,Kojo:2021ugu,Kojo:2021hqh,Iida:2022hyy}. 
The interplay between nuclear and quark equations of state is one of the central issues in neutron star phenomenology \cite{Drischler:2020fvz,Kojo:2020krb,Brandes:2022nxa,Gorda:2022jvk,Huang:2022mqp,Fujimoto:2022xhv,Marczenko:2022jhl}.

The above mentioned examples involve changes in effective degrees of freedom from composite to elementary particles.
In order to understand such transitions in QCD, it is crucial to study quark- and gluon-substructures of hadrons.
In this work we analyze the meson spectra within a schematic quark model, following the spirit and methodology of traditional constituent quark models \cite{DeRujula:1975qlm,Cornell,Godfrey:1985xj,Capstick:1986ter,Schnitzer:1978gq,Ebert:2009ub,Ebert:2009ua,Ebert:1997nk,Ebert:2002ig,Ebert:2011jc}
which have successfully reproduced not only the spectra but also the widths and hadronic couplings.
Our purpose here is to utilize some concepts and technology developed in the previous works and to present them in a schematic relativistic quark model.
Technical complications found in the previous works are considerably simplified by semi-phenomenological treatments that 
in turn enable us flexible approaches to more complicated systems.

While this work does not have much improvement in reproducing hadronic quantities compared to the previous works \cite{DeRujula:1975qlm,Cornell,Godfrey:1985xj,Capstick:1986ter,Schnitzer:1978gq,Ebert:2009ub,Ebert:1997nk,Ebert:2002ig,Ebert:2011jc},
this work has more focus on the application to hadronic matter in which the structure of hadrons may change.
In particular we manifestly present the sensitivity of hadronic spectra and structures to modeling in relativistic effects, short range correlations, and confining potentials. 
Such sensitivity may be obvious to experts working on quark models, 
but to the best of our knowledge is not much emphasized and is difficult to see from the outside of the community for the hadron spectroscopy.
Since the effective model parameters for quark dynamics in a hadron
may be influenced by hadron-hadron interactions,
it is important to specify which properties of hadrons can be stable until hadrons substantially overlap.

We analyze the global features of mesons from the light to heavy quark sectors, reproducing the spectra better than $10\%$ level.
We are particularly interested in the impacts of 
the relativistic kinematics \cite{Godfrey:1985xj,Capstick:1986ter}, 
scalar- vs vector-confinement \cite{Schnitzer:1978gq,Ebert:2009ub,Ebert:1997nk,Ebert:2002ig,Ebert:2011jc}, 
and semishort range correlations mediated by one-gluon exchanges.
All these effects are important especially in high density matter where particles become relativistic.
The distinction between scalar- and vector-confinement is connected to the relation between the chiral restoration and confinement, 
as the scalar vertices in vacuum is enhanced by the chiral symmetry breaking \cite{Alkofer:2006gz}.
The correlation between confinement and chiral symmetry breaking \cite{Casher:1979vw,Banks:1979yr,Coleman:1980mx}
is relevant to the chiral symmetry breaking/restoration 
in hadron-to-quark matter transitions \cite{Hatsuda:2006ps,Yamamoto:2007ah,Glozman:2007tv,Kojo:2009ha,Kojo:2011cn,Kojo:2010fe,Minamikawa:2021fln}.

As a quick application to hadronic matter under extreme conditions,
we consider a meson gas at finite temperature and examine the quark contents.
It has been known that a hadron resonance gas (HRG) model \cite{Venugopalan:1992hy,Karsch:2003vd,Karsch:2003zq}, 
where interactions among hadrons are neglected,
reproduces the results of lattice Monte-Carlo calculations up to $\sim 150$ MeV quite well.
In the context of HRG-QGP crossover, 
we are interested in how close quark contents in a HRG can be to those in a QGP.
The quark momentum distribution in a given meson is used to study the quark momentum distribution in a hot HRG.
A fuller examination of various quantities, e.g., chiral condensates, Polyakov loops, and so on, 
as well as the impacts of baryon resonances will be presented in separate publications.

This paper is organized as follows.
In Sec.~\ref{sec:RQM} we set up equations for bound state problems and explain how the relativistic kinematic factors are treated.
In Sec.~\ref{sec:V_eff} we discuss the potentials including various quark vertices.
In Sec.~\ref{sec:meson_spectra} we examine meson spectra and constrain our model parameters through fitting. 
In Sec. \ref{sec:Q_in} we calculate the single quark momentum distribution in a meson.
In Sec.~\ref{sec:HRG} we calculate the quark occupation probabilities in a meson gas.
Sec.~\ref{sec:summary} is devoted to summary. 

\section{Relativistic quark model}
\label{sec:RQM}

\subsection{The relations to the previous works}
\label{sec:prev_works}

As there are several successful constituent quark models which have strongly influenced our modeling,
we first comment on the similarity and difference between our work and the previous studies.

Our discussions on the relativistic kinematics (to be presented shortly in Sec.\ref{sec:basic_eq}) are largely motivated from the seminal works by Ebert et al. \cite{Ebert:2009ub,Ebert:1997nk,Ebert:2002ig,Ebert:2011jc}
which took into account the relativistic kinematics, the mixture of scalar-vector confining potentials, and short range interactions.
The model is able to reproduce the experimental spectra from light to bottom quark sectors quite well.
The authors replaced the quark energies appearing in various relativistic factors with some sort of energy constants 
which depend on the meson masses under discussions.
No expansion of quark momenta of $\sim \vp/m$ is used.
Our model applies similar simplifications to relativistic kinematic factors,
but our usage is more intuitive; we simply replace quark momenta with its average which are self-consistently estimated.
Our model adopts the mixture of scalar and vector confining potentials as in Refs.\cite{Ebert:2009ub,Ebert:1997nk,Ebert:2002ig,Ebert:2011jc},
but ours turn out to be dominated by the conventional scalar-type, while the latter by the vector-type.
The anomalous color magnetic moment used in Refs.\cite{Ebert:2009ub,Ebert:1997nk,Ebert:2002ig,Ebert:2011jc}, which substantially complicates the whole analyses,
is omitted in the present paper since we feel that introduction of such a term requires another justification and examination.

Another important work which we refer to is the work by Godfrey-Isgur for mesons \cite{Godfrey:1985xj} and its extension to baryons \cite{Capstick:1986ter}.
The authors treated the relativistic kinematics in a more complete manner than the averaging procedure in our modeling.
The potentials contain the scalar confining potential, short range Coulomb term, and various spin dependent potentials, 
all of which are convoluted with some phenomenological smearing functions including the quark mass dependence.
In our modeling, we do not directly use such smearing functions as we could not find simple reasonings for the parameterization.
But we do refer to the physical considerations in Refs.\cite{Godfrey:1985xj,Capstick:1986ter} when
we need to handle artificial singularities appearing 
in some short-range interactions.
As a whole our parametrization of various effects is based on more intuitive and straightforward considerations than Refs.\cite{Godfrey:1985xj,Capstick:1986ter}
and less sophisticated, but may be more flexible due to technical simplifcations.
Finally, Refs.\cite{Godfrey:1985xj,Capstick:1986ter} treat $\alpha_s$ as space-dependent to express its running behavior including the IR domain,
while we simply consider only typical values of $\alpha_s$ for given energy scales.
After getting the list of $\alpha_s$ for good fits, we extract plausible trends of $\alpha_s$ in the infrared.

In short, we aim at modeling which is intermediate between very popular non-relativistic quark models \cite{DeRujula:1975qlm,Cornell}
and more elaborated relativistic quark models in Refs.\cite{Godfrey:1985xj,Capstick:1986ter,Schnitzer:1978gq,Ebert:2009ub,Ebert:1997nk,Ebert:2002ig,Ebert:2011jc}.
We believe that treatments at the level of resolutions in this work are useful for grasping the qualitative features of many-body systems;
the advantage in our model is its intuitive feature which allows us to examine the correlation between the quark dynamics in a hadron and
the properties of matter.

\subsection{Basic equation}
\label{sec:basic_eq}

We compute a meson mass at rest and the corresponding wavefunction.
We solve equations for relativistic quarks.
Our starting point is ($\int_{\bm q} \equiv \int \rmd^3  {\bm q}/(2\pi)^3$)
\beq
\big[ M - E_1 (\vp) - E_2 (\vp) \big] \Psi (\vp)
= \int_{\vq} V (\vp,  {\bm q}) \Psi ( {\bm q}) \,,
\label{eq:sch_eq0}
\eeq
where $M$ is the meson mass, $E_{i=1,2} (\vp) = \sqrt{ \vp^2+m_{i}^2 }$ the energies of quarks 1 and 2, and $\Psi (\vp)$ the meson wavefunction at rest.
The two body effective potential $V (\vp,  {\bm q})$ includes the Dirac spinors and the matrix elements at the vertices.
We take the center of mass frame of the two quarks, and $\vp$, ${\bm q}$ are relative momenta (here the quark momenta are $\vp_1= - \vp_2=\vp$ or ${\bm q}$).
$m_{i=1,2}$ are the constituent quark masses of $\sim 0.3$ GeV for light quarks, $\sim 0.5$ GeV for strange quarks, and $\sim 1.5$ GeV for charm quarks.

Following the approaches in Refs.\cite{Ebert:2009ub,Ebert:1997nk,Ebert:2002ig,Ebert:2011jc}, 
we rewrite Eq.(\ref{eq:sch_eq0}) into a Schr{\"o}dinger-type differential equation. 
This is possible even without applying the nonrelativistic expansion. 
We first multiply a factor
\beq
I_M (\vp) = \frac{\,  \big[ M+E_1+E_2 \big] \big[ M^2 - (E_1 - E_2)^2 \big] \,}{ 4M^2 } \,,
\label{eq:I_M}
\eeq
to Eq.(\ref{eq:sch_eq0}), and obtain a Sch{\"o}dinger-like equation
\beq
\big( b_M - \vp^2 \big) \Psi (\vp)
= \int_{ {\bm q}} \calV_M (\vp, {\bm q}) \Psi ( {\bm q}) \,,
\label{eq:sch_eq1}
\eeq
where we wrote 
\beq
\calV_M (\vp, {\bm q}) \equiv I_M (\vp) V (\vp, {\bm q})\,,
\eeq
and
\beq
b_M = \frac{\, \big[ M^2 - (m_1+m_2)^2 \big]  \big[ M^2 - (m_1-m_2)^2 \big] \,}{4M^2 } \,,
\label{eq:b_M}
\eeq
is a constant depending on the meson mass $M$.

Although the equation looks a usual eigenvalue equation, the actual determination of the meson mass $M$ is more complex because $M$ appears in kinematic factors in $\calV_M$.
To determine $M$, we find the eigenvalue $\lambda_M$ of an equation $ \big( \vp^2 + \calV_M \big) |\Psi\ra = \lambda_M | \Psi \ra$ for a given $M$, and check whether $\lambda_M$ coincides with $b_M$.
Only the value $M$ satisfying $\lambda_M = b_M$ is adopted as a physical meson spectrum.


\subsection{Averaging kinematic factors}
\label{sec:ave_kin}

\begin{figure}[bt]
\vspace{-0.8cm}
\begin{center}	
	\includegraphics[width=4.0cm]{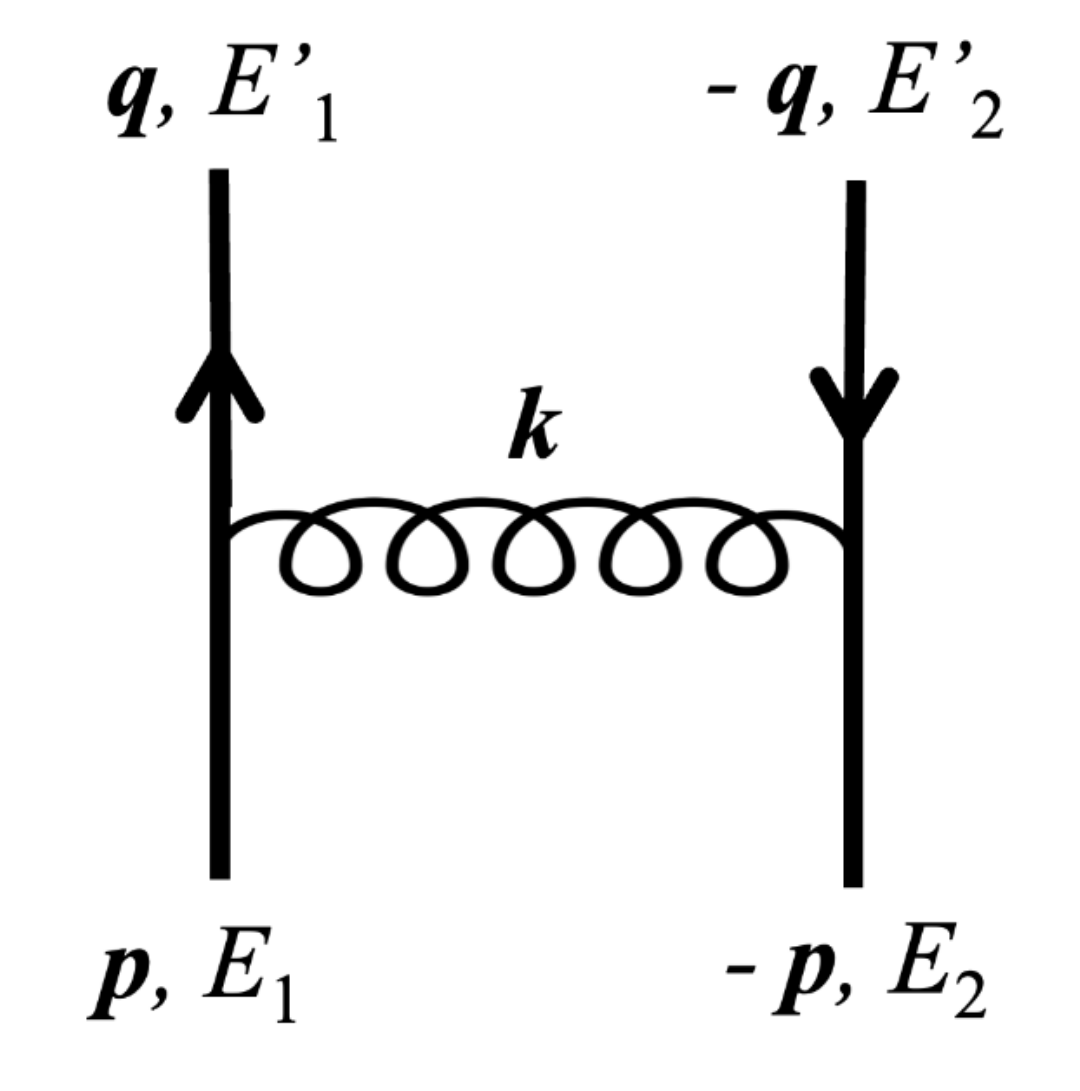}
		\vspace{-0.8cm}
\caption{  A graph for two-body scattering with the notations corresponding to those in the main text.
		 }
		 		\vspace{-0.8cm}
		 \label{fig:graph_OGE}
\end{center}		
		
\end{figure}

The effective potentials $\calV_M(\vp, {\bm q})$ in Eq.(\ref{eq:sch_eq1}) contain various quark kinematic factors associated with the energies and spinors.
In particular quark momenta can appear in denominators.
To simplify the calculations we replace some of momenta with the average values for a given meson.
First we rewrite $\calV_M (\vp,  {\bm q})$ as a function of $E_{1,2} (\vp)$ and $E'_{1,2} ( {\bm q})$, 
the average momentum $\vP$ before and after interactions (not total momentum!), and the momentum transfer $\vk$ as
\beq
\vP = \frac{\, \vp +  {\bm q} \,}{2}\,,~~~ \vk = \vp -  {\bm q} \,.
\eeq
The potential can be expressed as 
\beq
\calV_M(\vp, {\bm q}) = \tilde{\calV}_M(E_{1,2}, E'_{1,2}; \vP, \vk ) \,.
\eeq
Here, for later convenience we wrote $E,E'$ as if they are independent of $\vP, \vk$. 

Our approximation takes the time average of $\vP^2$ which has a typical value for a given meson.
We approximate the energies as
\beq
E_i (\vp), E'_i ( {\bm q}) ~\rightarrow ~ \bar{E_i} = \sqrt{ m_i^2 + \bar{\vP}^2 } \,,
\eeq
where $\bar{\vP}^2$ is the average of $\vP^2$.
Next, if we encounter expressions such as $P_i$ or $P_i P_j$, we take the average as
\beq
P_i ~\rightarrow~ 0\,,~~~ P_i P_j ~\rightarrow \delta_{ij} \bar{\vP}^2/3 \,.
\label{eq:naive_average}
\eeq
and so on. With these procedures 
\beq
\calV_M(\vp, {\bm q}) 
~\rightarrow~ 
\overline{\calV}_{M, \bar{\vP}^2} ( \vk)
 \,.
\eeq
For this effective potential, the only variable is $\vk$, and one can take the Fourier transform of Eq.(\ref{eq:sch_eq1}) to obtain
\beq
b_M  \Psi (\vecr)
= \big[\, - \nabla^2 + \overline{\calV}_{M, \bar{\vP}^2} (\vecr) \, \big] \Psi (\vecr) \,.
\label{eq:sch_eq1}
\eeq
In practice, we substitute some values of $\bar{\vP}^2$, find the eigenstates, and compute\footnote{In practice we compute $\la - \nabla^2 \ra = \la (b_M-\bar{\calV}_M) \ra$ for $\la \bar{\vP}^2 \ra$. } 
the expectation value $\la \vP^2 \ra$ to check whether our choice of  $\bar{\vP}^2$ coincides with $\la \vP^2 \ra$.
This forms a self-consistent equation for a given $M$ (which does not necessarily give the solution, $\lambda_M = b_M$).
In this work we varied $\bar{\vP}^2$ with only grids of $0.01\, {\rm GeV}^2$ without demanding perfect consistency.
The choice of $\bar{\vP}^2$ depends on the potential and the state we are looking at.
For example, if the potential for the equal quark mass case is the pure Coulomb type,\footnote{We balance $\Delta p^2/2m$ and $\alpha_s/\Delta r \sim \alpha_s \Delta p$ to get $\Delta p \sim \alpha_s m$. Note that a large $m$ does not necessarily justifies a non-relativistic treatment; for a large $\alpha_s$, the strong attraction leads to deeply bound states or very compact objects in which kinetic energies and momenta are large, requiring relativistic treatments. } 
then $\bar{\vP}^2\sim ( \alpha_s m)^2$, with which the non-relativistic approximation is valid only for $\alpha_s(m^2) \ll 1$. 
The linear rising potential modifies this simple rule, but in general we found $|\vP|$ 
is the order of $\gtrsim 0.3$ GeV $\sim \alpha_s(m_{u,d}^2) m_{u,d}$ for light quarks and $\gtrsim 0.5$ GeV $\sim \alpha_s (m_c^2) m_c$  for charm quarks.

There is a qualification on the above averaging procedures Eq.(\ref{eq:naive_average}) for $\vP$. 
We actually have dropped the appearance of some angular momentum operators which otherwise leave the $\vL\cdot {\bm S}$ type operators. 
To correctly keep such terms, we add an extra rule here.
When $\vP$ appears in the combination of $\vk \times \vP$ together with the wavefunction $\Psi({\bm q})$, 
we use $\vP = {\bm q} +\vk/2$ to rewrite $\vk \times \vP = \vk \times {\bm q}$. 
Taking the Fourier transform,
\beq
\int_{\vp, {\bm q}} \rme^{\rmi \vp \cdot \vecr} ( \vk \times {\bm q} ) f(\vk) \Psi({\bm q})
= - {\bm \nabla} f(\vecr)  \times  {\bm \nabla} \Psi(\vecr)  \,. 
\eeq
For a rotationally symmetric function $f(\vecr) = f(r)$,
\beq
{\bm \nabla} f(\vecr)  \times  {\bm \nabla} \Psi(\vecr) 
= \frac{\, \partial_r f(r)  \,}{\, r \,} \vL \Psi(\vecr) \,,  
\eeq
with $\vecr \times {\bm q} = \vecr_1 \times {\bm q} + \vecr_2 \times ( -{\bm q} ) = \vl_1 + \vl_2 = \vL$ 
being the total angular momentum operator acting on $\Psi$.

\subsection{Several kinematic limits}
\label{sec:kin_limit}

Here we analyze various kinematic limits. 
We give a brief summary here for the factors given in Eqs.(\ref{eq:I_M}), (\ref{eq:b_M}), and the eigvenvalue equation Eq.(\ref{eq:sch_eq1}).

For a heavy-heavy meson with $m_1 = m_2 = m_h$, the kinematic factor is $(M \sim 2m_h)$
\beq
I_M^{ hh} 
\simeq  \frac{\, M + 2 m_h \,}{ 4} 
\simeq m_h \,,
\label{eq:I_Mhh}
\eeq
and
\beq
b_M^{hh} 
= M - 2m_h \,.
\label{eq:b_Mhh}
\eeq
With the potential $\bar{V}^{hh} = \bar{V}_{m_1,m_2\rightarrow \infty}$ for this limit,
our bound state problem then becomes
\beq
\big( M - 2m_h \big) |\Psi\ra 
= \bigg( \frac{\, \vp^2 \,}{\, m_h \,} + \bar{V}^{ hh} \bigg) |\Psi\ra \,,
\label{eq:Vhh}
\eeq
%
where $M-2m_h$ is a binding energy.
Here $M$ only appears as an eigenvalue of the equation; the kinematic factor and potentials are $M$-independent
as in usual non-relativistic problem.

For a heavy-light meson with $m_1 = m_h \gg m_2 = m_l $, the kinematic factor is $(M \sim m_h)$
\beq
I_M^{hl}  (\vp) 
\simeq
M + \bar{E}_l - m_h  \,,
\label{eq:I_Mlh}
\eeq
%
where $\bar{E}_l = \sqrt{ \bar{\vP}^2 + m_l^2 }$,
and
\beq
b_M^{hl} 
= \big[ M - m_h - m_l \big]  \big[ M - m_h + m_l \big]  
 \,.
\label{eq:b_Mlh}
\eeq
Our bound state problem then becomes
%
\beq
&& \big[ M - m_h - m_l \big] \frac{\,  M -m_h + m_l \,}{\, M - m_h + \bar{E}_l \,}  |\Psi\ra 
\nonumber \\
&& = \bigg( \frac{\, \vp^2 \,}{\, M -m_h + \bar{E}_l \,} + \bar{V}^{ hl} \bigg) |\Psi\ra \,.
\eeq
where $\bar{V}^{hl} = \bar{V}_{m_1 \rightarrow \infty}$
We note that the eigenvalue problem is more complex than the heavy-heavy systems 
due to the appearance $M$ in the kinetic term.
We note that $M-m_h$ can be comparable to $m_l$ and $\bar{E}_l$.

Finally we consider the relativistic limit where $M \sim 2 |\vp| \gg m_1, m_2$,
\beq
I_M^{R} (\vp) 
\simeq \frac{\, M+ 2 |\vp| \,}{ 4 } \sim M \,,
~~~~
b_M^{R} (\vp) 
\simeq \frac{\, M^2 \,}{ 4 } \,,
\label{eq:I_M_R}
\eeq
with which 
\beq
M  |\Psi\ra 
= \big(\, 2 |\vp| + \bar{V} \,\big) |\Psi\ra \,.
\eeq
This limiting behavior is important for highly excited mesons.
For example, mesons in the Regge trajectories have momenta $ |\vp| \sim \sqrt{n_r \lqcd}$ for radial excitations and $\sim \sqrt{l \lqcd}$ for the angular excitations.

\section{ potentials}
\label{sec:V_eff}

The potential $V(\vp,{\bm q})$, including the quark vertices, is given by the sum of various components which generally take the forms
\beq
&& V_n (\vp, {\bm q})_{s_1 s_2}^{s'_1 s'_2} 
\nonumber \\
&& = \big[ \bar{u}^{s'_1}_1(\vp) u^{s_1}_1 ({\bm q}) \big]
~ U_n (\vk)_{s_1 s_2}^{s'_1 s'_2} ~
\big[ \bar{u}^{s'_2}_2(-\vp) u^{s_2}_2 (-{\bm q}) \big]\, ,
\eeq
where $s_1, s_2,...$ are the spin indices.
For $U_n$, we consider the Lorentz scalar- and vector-confining potentials, and semishort range effects from one gluon exchanges.
The spinor takes the form
\beq
\hspace{-0.5cm}
u_i^s (\vp) 
= \calN
 \bigg(
\begin{matrix}
1 \\
\frac{ {\bm \sigma} \cdot \vp }{\, E_i (\vp) + m_i \,}
\end{matrix}
\bigg) \chi^s \,,
~~
 \calN 
= \sqrt{ \frac{\, E_i (\vp) + m_i \,}{ 2E_i (\vp) }  } \,,
\eeq 
with $\chi^{ \up} = (1,0)^T$ and $\chi^{ \down} = (0,1)^T$.

\subsection{Scalar and vector confining potentials}
\label{sec:scalar_vector_confining}

We use the scalar confining potential (after taking the Fourier transform)
\beq
 U_{\rm conf}^S (\vecr)_{s_1 s_2}^{s'_1 s'_2} 
 = f_S\, \delta^{s'_1}_{s_1 } \delta^{s'_2}_{ s_2} \times \sigma r \,.
\eeq
and the vector confining potential (metric: $\eta^{\mu \nu} = \eta_{\mu \nu} = {\rm diag} (1, -1,-1,-1)$)
\beq
 U_{\rm conf}^V (\vecr)_{s_1 s_2}^{s'_1 s'_2} 
 = (1-f_S)\, \eta^{\mu \nu} (\gamma_\mu)^{s'_1}_{s_1 } (\gamma_\nu)^{s'_2}_{ s_2}  \times \sigma r \,.
\eeq
The parameter $f_S$ represents the relative strength of the scalar and vector parts.
In the heavy quark limit, $\bar{u} u \rightarrow 1$, $\bar{u} \gamma_0 u \rightarrow 1$, and $\bar{u} \gamma_i u \rightarrow 0$. 
The choice of $f_S$ does not matter in the heavy quark limit; 
only the sum of the scalar and the zeroth component of the vector potentials appears, leaving the linear rising potential $\sigma r$.
We choose the standard value, $\sigma \simeq 0.18\, {\rm GeV}^2 \simeq 0.91\, {\rm GeV/fm}$ extracted 
from heavy quark spectra and Wilson loops in the lattice QCD (for a pedagogical review, see e.g. Ref.\cite{Bali:2000gf}).
Meanwhile for light quark sectors there are corrections of $\sim \vP^2/E^2$ which come from the kinematics
and the spatial components of the vector vertices, see Eq.(\ref{eq:V_c0}) in Sec.\ref{sec:full_expressions}.
The effective string tension away from the heavy quark limit has been studied in the Bethe-Salpeter approach in lattice QCD \cite{Kawanai:2011xb,Kawanai:2015tga,Nochi:2016wqg}.
It has been seen that, as quark masses decrease to $\sim 1$ GeV, the effective string tensions initially decrease to charm quark region by $\sim 10\%$, and then increase toward the lighter quark region  \cite{Kawanai:2011xb}. 
As we will see, it is necessary, for each channel, to tune the value for $f_S$ by $\lesssim 10\%$ accuracy to reproduce radial and orbital excitations. 
We examine the trend of $f_S$ from light to heavy quark sectors, and will find $f_S$ to gently grow toward heavier quark systems.

\subsection{Semishort range correlations}
\label{sec:semishort_range}

While the confining potential is crucial to describe the excited states,
the low-lying states with small radial and angular excitations are sensitive to the semi-short\footnote{We attach ``semi" to reserve ``short range" for the perturbative regime.} range correlations.
The standard description uses the one gluon exchange (OGE) \cite{DeRujula:1975qlm,Cornell,Godfrey:1985xj,Capstick:1986ter}
\beq
U_{\rm OGE}^V (\vk)_{s_1 s_2}^{s'_1 s'_2} 
= - \frac{\, 4 \,}{3}\, \alpha_s D_{\mu \nu}(\vk)  (\gamma^\mu)^{s'_1}_{s_1 } (\gamma^\nu)^{s'_2}_{ s_2} \,,
\eeq
with the Coulomb gauge form,
\beq
D_{00} (\vk) =  \frac{\,  4\pi \,}{\vk^2}  \,, ~~~~~~ D_{0i} (\vk) = D_{i0} (\vk) = 0 \,, \nonumber \\
D_{ij} (\vk) =  \frac{\,  4\pi \,}{ \vk^2 } \bigg( - \delta_{ij} + \frac{\, k_i k_j \,}{\vk^2} \bigg) \,.~~~~~~~
\eeq
The zeroth component produces the color electric potential at short distance which appears from the light to heavy quark sectors.
Meanwhile, the spatial components couple to the vertices $\bar{u} \gamma_i u $ whose magnitudes are proportional to the velocities of quarks, $\sim \vp/E$.
The latter is particularly important for relativistic regimes.

\begin{figure}[bt]
\vspace{-0.cm}
\begin{center}	
	\includegraphics[width=8.5cm]{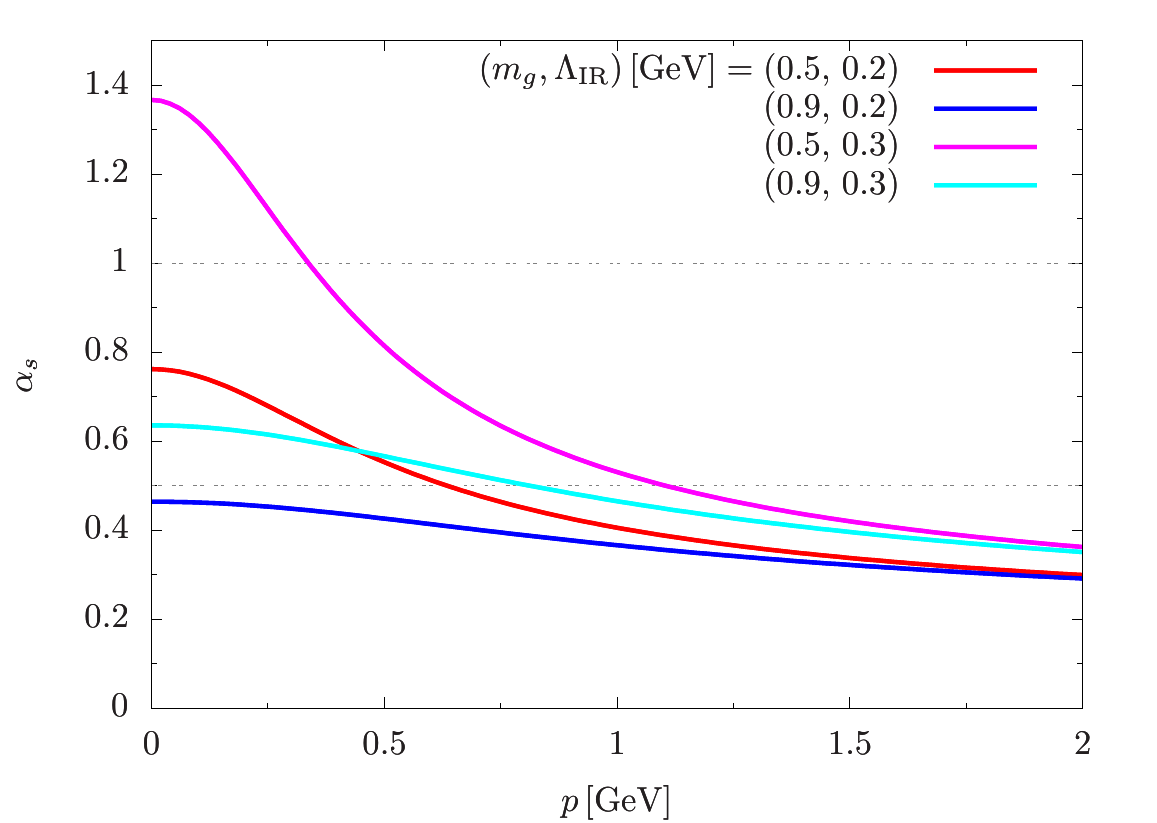}

		\vspace{-0.2cm}
\caption{  $\alpha_s$ in Eq.(\ref{eq:alpha_s_para}) as a function of (Euclidean) momentum $p \equiv \sqrt{Q^2}$ for several combinations of $m_g$ and $\Lambda_{\rm IR}$.
		 }
		 		\vspace{-0.4cm}
		 \label{fig:alpha_freezing}
\end{center}		
		
\end{figure}

We use the OGE for the momentum transfer of $\lesssim 1$ GeV and hence have to specify how $\alpha_s$ behaves in the non-perturbative domain.
How to define $\alpha_s$ is a highly nontrivial issue in its own (for a comprehensive summary, see Ref. \cite{Deur:2016tte}).
Our reference is the IR freezing picture for the running coupling constant \cite{Deur:2016tte}
\beq
\alpha_s (Q^2) = \frac{\, 4\pi \,}{\, \beta_0 \ln \frac{\, Q^2 + m_g^2 \,}{\, \Lambda_{\rm IR}^2 \,} } \,,
\label{eq:alpha_s_para}
\eeq
where $\beta_0= (11\Nc-2\Nf)/3$ with $\Nc$ and $\Nf$ being the number of colors and flavors, $Q^2$ a squared momentum transfer in the euclidean signature, $m_g$ a parameter of $\sim 0.5$-0.9 GeV, and $\Lambda_{\rm IR} \simeq 0.2$-0.3 GeV.
The behaviors $\alpha_s$ are shown in Fig.\ref{fig:alpha_freezing} where substantial dependence on $m_g$ and $\Lambda_{\rm IR}$ can be seen at $p \lesssim 1$ GeV.
The inclusion of $m_g$ is motivated by the observation that gluonic fluctuations at long wavelength cannot grow arbitrarily in amplitudes
and hence temper the RG evolution toward the IR limit, removing the perturbative Landau pole.
The IR finite gluonic fluctuations are either due to the confining effects or the gluon saturation due to the self-coupling \cite{Gribov:1977wm}.
Models of massive gluons also lead to similar descriptions \cite{Curci:1976bt,Mandula:1987rh,Reinosa:2017qtf,Reinosa:2013twa,Suenaga:2019jjv,Song:2019qoh,Kojo:2021knn}.

For the choice of $Q^2$, its typical magnitude is correlated with the typical momenta of quarks in a given problem, so we suppose $Q^2\sim \bar{\vP^2}$ or $\bar{E}^2$.
But in practice we use Eq.(\ref{eq:alpha_s_para}) as a mere qualitative guide, treating $\alpha_s$ as one of fitting parameters for mesons in a given channel, see Tables in Sec.\ref{sec:meson_spectra}.
Then we examine a general tendency of $\alpha_s$ {\it posteriori}.
It turns out that $\alpha_s$ is $\sim 0.7$-$0.8$ for light and strange quarks, and $\sim 0.4$ for charmed quarks, reasonably consistent with the behaviors shown in Fig.\ref{fig:alpha_freezing}.

\subsection{Relativistic vertices}
\label{sec:rela_vertices}

Computing the quark vertices coupled to gluons, we encounter the following types of terms
\beq
\sim \frac{\, k_i \,}{\, E_1 (\vP+\vk/2) \,} U(\vk) \frac{\, k_j \,}{\, E_2 (\vP-\vk/2) \,} \,,
\eeq
where the factor $\sim \vk/E$ comes from the velocity factors at vertices.
In our framework $E_{1,2}$ are replaced with the average $\bar{E}_{1,2}$.
This simplification would leave artifacts as discussed below.

Neglecting the $\vk$-dependence in the denominators, we obtain the factors $\sim k_i k_j/\bar{E}_1 \bar{E}_2$.
Such high powers in $\vk$ combined with $U(\vk)$ often result in singular potentials.
For instance, with $i=j$ and $U(\vk)=1/\vk^2$, we have
\beq
\sim \frac{\, \vk^2 \,}{\, \bar{E}_1 \bar{E}_2 \,} \frac{\, 1 \,}{\, \vk^2 \,}
~\rightarrow~  \frac{\, 1 \,}{\, \bar{E}_1 \bar{E}_2 \,} \delta(\vecr) \,,
\eeq
after taking the Fourier transform. This singular behavior is not problematic as far as we evaluate it in perturbation theories, but it would become problematic in unperturbed treatments. The singular behavior is just an artifact of abusing the replacement $|\vk|/E\rightarrow |\vk|/\bar{E}$ at large momenta; instead the correct behavior is $|\vk|/E\rightarrow 1$ in large $\vk$ limit and becomes harmless.

Taking the above considerations into account, we write the spinor matrices as
\beq
\bar{u}_i (\vp) u_i (\vq)
&\rightarrow& \frac{\, m_i \,}{\, \bar{E}_i \,} - \frac{\, (\rmi \vk)^2 {\mathcal K}_{i}  + 4 {\bm \sigma}_i \cdot (\rmi \vk \times \vP) {\mathcal L}_{i}  \,}{\, 8\bar{E}_i \zeta_i \,} \,,
\nonumber \\
\bar{u}_i (\vp) \gamma_0 u_i (\vq)
&\rightarrow& 1 + \frac{\, (\rmi \vk)^2 {\mathcal K}_{i}  + 4 {\bm \sigma}_i \cdot (\rmi \vk \times \vP) {\mathcal L}_{i} \,}{\, 8\bar{E}_i \zeta_i \,} \,,
\nonumber \\
\bar{u}_i (\vp) {\bm \gamma } u_i (\vq)
&\rightarrow&  \frac{\, 2 \vP - ( \rmi \vk \times  {\bm \sigma}_i ) {\mathcal J}_{i} \,}{\, 2\bar{E}_i  \,} \,.
\label{eq:k_E_replacement}
\eeq
where ${\mathcal K}_i, {\mathcal L}_i$ and ${\mathcal J}_i$ ($i=1,2$) are form factors as functions of $\vk$ to cutoff the UV singularity associated with large $| \vk |$.
If we set ${\mathcal K}_i = {\mathcal L}_i = {\mathcal J}_i =1$ we recover expressions without cutoff effects.  
In practice we assume the Gaussian factors,
${\mathcal K}_i , {\mathcal L}_i , {\mathcal J}_i \sim \rme^{- {\rm const.} \times \vk^2/\bar{E}_i^2 }$, with which we get the expression similar to nonrelativistic expression at low $\vk$.
We distinguish the coefficients ${\mathcal K}_i , {\mathcal L}_i , {\mathcal J}_i$ as they characterize different physics.

We proceed with this approximation and take the Fourier transform to get the coordinate space expression.
But we further simplify the expression while keeping the above qualitative features. The steps are detailed in Appendix.\ref{app:form_factors}.
In short, we include the form factors for each potential and take the Fourier transform.
For instance, terms with form factors ${\mathcal K}_1$ and ${\mathcal K}_2$ are converted into
\beq
\frac{\, (\rmi k_i) (\rmi k_j )  \,}{\, f( E_1,E_2)  \,} {\mathcal K}_1 {\mathcal K}_2U(\vk) 
& ~\rightarrow~ &
\frac{\, \partial_i \partial_j  \,}{\, f ( \bar{E}_1, \bar{E}_2) \,} U_{\rm reg}^{ {\mathcal K}_1  {\mathcal K}_2 } (r)  \,,
\eeq
where $f (E_1,E_2)$ is a general function of $E_{1,2}$ appearing in the denominator,
and $U_{\rm reg}^{ {\mathcal K}_1  {\mathcal K}_2 } $, with the upperscripts specifying the types of form factors, 
represents the short distance contributions,
\beq
U_{\rm reg}^{ {\mathcal K}_1  {\mathcal K}_2 }  
= \rme^{- \beta_{ {\mathcal K}_1  {\mathcal K}_2 } ^2 r^2} \times U^{ {\mathcal K}_1  {\mathcal K}_2 } _{\rm sh}  \,,
\label{eq:def_Uinter}
\eeq
with the strength averaged over distance $\sim \beta^{-1}$,
\beq
\hspace{-0.5cm}
&&
U_{\rm sh}^{ {\mathcal K}_1  {\mathcal K}_2 } 
= 
\int_0^\infty \! \rmd \xi \, \xi^2 w_{ {\mathcal K}_1  {\mathcal K}_2 } (\xi) U^{ {\mathcal K}_1  {\mathcal K}_2 } (\xi) \,,
~~ 
\nonumber \\
&&
 w_{ {\mathcal K}_1  {\mathcal K}_2 } (\xi) = \frac{\, 4 \beta_{ {\mathcal K}_1  {\mathcal K}_2 } ^3 \,}{\, \sqrt{\pi } \,} \, \rme^{-\beta_{ {\mathcal K}_1  {\mathcal K}_2 }^2 \xi^2} \,.
\eeq
The parameter $\beta \sim \bar{E} $ is related to our form factors and hence depend on the types of potentials.
In general, we write, for $s = \calK_{i}, \calL_i, \calK_1\calK_2,  \calJ_1 \calJ_2$,
\beq
&&U_{\rm reg}^{ s }  
= \rme^{- \beta_{ s } ^2 r^2} \times U^{ s } _{\rm sh}  \,,~~~
\nonumber \\
&&
\hspace{-0.5cm}
U_{\rm sh}^{ s } 
= 
\int_0^\infty \! \rmd \xi \, \xi^2 w_{ s} (\xi) U^{ s } (\xi) \,,
~~ 
w_{ s } (\xi) = \frac{\, 4 \beta_{ s } ^3 \,}{\, \sqrt{\pi } \,} \, \rme^{-\beta_{ s}^2 \xi^2}
 \,.
\eeq
Here $\beta_s$'s are parametrized as
\beq
&& \beta_{\calK_{i}} = C_{\calK_{i}} \bar{E}_{i}\,,
~~~~\beta_{\calL_{i}} = C_{\calL_{i}} \bar{E}_{i}\,, 
\nonumber \\
&& \beta_{\calK_{1} \calK_{2}} = C_{\calK_{1} \calK_{2}} \bar{E}_{\rm red} \,,
~~~~\beta_{\calJ_{1} \calJ_{2}} = C_{\calJ_{1} \calJ_{2}} \bar{E}_{\rm red} \,,
\eeq
where $C$'s are treated as constants of $O(1)$, and
\beq
\bar{E}_{\rm red}^{-2} \equiv \bar{E}_1^{-2} + \bar{E}_2^{-2} \,.
\eeq
Below we often omit the subscripts ${\mathcal K}_i, {\mathcal L}_i$, and ${\mathcal J}_i$ to avoid busy notations.

In this paper we work only up to the second order of $\vk^2$ to which each vertex contribute to one power of $k$.
We use the following notations for our replacement procedures,
\beq
 (\rmi \vk)_i (\rmi \vk)_j \tilde{U}   ~&\rightarrow&~ 
 \delta_{ij} \bar{U}_2 (r)/3 + Q_{ij}  \bar{U}_{2T} (r)/3  \,,
 \nonumber \\
 ( \rmi \vk \times \vP) \tilde{U} ~&\rightarrow&~ \vL \bar{U}_{1L}  (r)\,, 
 \nonumber \\
  ( \rmi \vk \times \vP)_i   ( \rmi \vk \times \vP)_j \tilde{U} 
 ~&\rightarrow&~ \calP_{ij}  \bar{U}_{2\calP} (r)+ L_i L_j \bar{U}_{2L^2}(r) \,, 
  \nonumber \\
\eeq
where we have defined 
\beq
Q_{ij} = \delta_{ij} - \frac{\, 3 x_i x_j \,}{\, r^2 \,} \,,~~~
\calP_{ij} = \delta_{ij} \vP^2 - P_i P_j \,.
\eeq
The subscript of $\bar{U}$ indicates the powers of $\vk$.
Explicitly, each potential is computed from $U_{\rm inter}$ as
\beq
&&
\bar{U}_2  = {\bm \nabla}^2 U_{\rm reg}\,,~~~~
\bar{U}_{2 T} 
= - \bigg(  \frac{\, \rmd^2  \,}{\, \rmd r^2 \,} - \frac{1}{\, r \,} \frac{\, \rmd \,}{\, \rmd r \,} \bigg)U_{\rm reg} \,,
\nonumber \\
&& ~~~~~~~~~
\bar{U}_{1L} (r) = \bar{U}_{2\calP} (r) 
= \frac{1}{\, r \,} \frac{\, \rmd U_{\rm reg} \,}{\, \rmd r \,} \,, 
\nonumber \\
&& ~~~~~~~
\bar{U}_{2 L^2} (r) 
= \frac{1}{\, r^2 \,} \bigg( \frac{\, \rmd^2  \,}{\, \rmd r^2 \,} -  \frac{1}{\, r \,} \frac{\, \rmd \,}{\, \rmd r \,} \bigg)U_{\rm reg}
\,.
\eeq
Some useful formulae are summarized in Appendix.\ref{app:pot_deri}.

\subsection{The full expressions of potentials}
\label{sec:full_expressions}

Convoluting the quark vertices and the potentials $U$'s, we obtain
\beq
\bar{V} 
&=&
 \bar{V}_{ C }^{(0)} + \bar{V}_{ C }^{(2)} + \bar{V}_{ ss}  ( {\bm \sigma}_1 \cdot {\bm \sigma}_2 )
+ \bar{V}_{ L^2 }  {\bm L}^2 
\nonumber \\
&& +
 \sum_{i=1,2} \bar{V}^{(i)}_{ sL } ( {\bm \sigma}_i \cdot \vL )
+ \bar{V}_{ (sL)^2 } ( {\bm \sigma}_1 \cdot \vL )( {\bm \sigma}_2 \cdot \vL )
\nonumber \\
&& + \,  \bar{V}_{T} \big( {\bm \sigma_1}  \cdot {\bm Q} \cdot  {\bm \sigma_2} \big) \,.
\label{eq:full_expressions_pot}
\eeq
The first term is the central potential without spatial derivatives
\beq
\bar{V}_{ C }^{(0)}
 = 
\bigg( 
1 + \frac{\,  \bar{\vP}^2  \,}{\, \bar{E}_1  \bar{E}_2   \,}   \bigg) 
 \bigg( U^V_{\rm conf} +U_{\rm OGE}^V \bigg)
+ 
 \frac{\, m_1 m_2 \,}{\, \bar{E}_1 \bar{E}_2 \,} U^S_{\rm conf} 
\,.
\nonumber \\
\label{eq:V_c0}
\eeq
As we will see, this term plays the dominant roles in determining the overall spectra.
All the other terms contain the spatial derivatives of the potentials.
The central potential with spatial derivatives is $(\zeta_i \equiv \bar{E}_i + m_i$)
\beq
\hspace{-0.5cm}
\bar{V}_{ C }^{(2)}
&=& 
\frac{\, 1  \,}{\, 8 \,} 
\bigg(
 \frac{\, \bar{U}^{ {\mathcal K}_1 }_2 \,}{\,  \bar{E}_1 \zeta_1  \,} 
+
 \frac{\, \bar{U}^{ {\mathcal K}_2 }_2  \,}{\,   \bar{E}_2 \zeta_2  \,} 
\bigg)
 \bigg|_{U= U_{\rm OGE}^V + U_{\rm conf}^V }
\nonumber \\
&& - 
\frac{\, 1  \,}{\, 8 \,} 
\bigg(
 \frac{\, \bar{U}^{ {\mathcal K}_1 }_2 \,}{\,  \bar{E}_1 \zeta_1  \,}  \frac{\, m_2 \,}{\,  \bar{E}_2  \,} 
+
 \frac{\, \bar{U}^{ {\mathcal K}_2 }_2  \,}{\,   \bar{E}_2 \zeta_2  \,}   \frac{\, m_1 \,}{\,  \bar{E}_1  \,} 
\bigg)
\bigg|_{U= U_{\rm conf}^S } \,.
\eeq
The spin-spin interaction is given by
\beq
\bar{V}_{ ss}  
& = &
\frac{\, 1 \,}{\, 6 \bar{E}_1 \bar{E}_2  \,} \bigg( \bar{U}_{2}^{ {\mathcal J}_1  {\mathcal J}_2 }   
	+ \frac{\, \vP^2 \,}{\, \zeta_1 \zeta_2 \,}  \bar{U}_{2\calP}^{ {\mathcal J}_1  {\mathcal J}_2 }        \bigg) \bigg|_{U= U_{\rm OGE}^V + U_{\rm conf}^V}
	\nonumber \\
& & +
\frac{\, \bar{\vP}^2 \,}{\, 6 \bar{E}_1 \bar{E}_2 \zeta_1 \zeta_2  \,} \bar{ U}_{2\calP}^{ {\mathcal J}_1  {\mathcal J}_2 }     \bigg|_{U= U_{\rm conf}^S } \,.
\eeq
The potential giving additional centrifugal forces
\beq
\bar{V}_{ {\bm L^2} }  
=   	
\frac{\, 1 \,}{\, 2 \bar{E}_1 \bar{E}_2 \,} \bar{U}_{1L}^{ {\mathcal J}_1  {\mathcal J}_2 } \big|_{U= U_{\rm OGE}^V } \,,
\eeq
which arise only from the $k_i k_j/\vk^2$ term in the one gluon exchange.
The potential for the spin-orbit coupling is given by
\beq
\bar{V}^{(i=1)}_{ sL } 
&=& \bigg( \frac{1}{  \bar{E}_1 \zeta_1} + \frac{1}{\, \bar{E}_1  \bar{E}_2 \,} \bigg) \frac{\, \bar{U}^{ {\mathcal L}_1 }_{1L}  }{2} \bigg|_{U= U_{\rm OGE}^V + U_{\rm conf}^V } 
\nonumber \\
&&- ~  \frac{1}{\, \bar{E}_1 \zeta_1 \,}  \frac{m_2}{\, \bar{E}_2 \,}  \frac{\, \bar{U}^{ {\mathcal L}_1 }_{1L}  }{2} \bigg|_{U= U_{\rm conf}^S } \,.
\eeq
The $ \bar{V}^{(i=2)}_{ sL }$ is obtained by swapping 1 and 2.

The potential for the square of the spin-orbit coupling operators is
\beq
\bar{V}_{ (sL)^2 }
= \frac{\, \bar{U}_{2L^2}^{ {\mathcal J}_1 {\mathcal J}_2 } \,}{\,  4  \bar{E}_1  \bar{E}_2 \zeta_1 \zeta_2 \,} \bigg|_{U= U_{\rm OGE}^V + U_{\rm conf}^V + U_{\rm conf}^S } \,.
\eeq
Finally, the tensor potential is
\beq
\bar{V}_T
= - \frac{\,  \bar{U}_{2T}^{ {\mathcal J}_1 {\mathcal J}_2 } \,}{\, 12  \bar{E}_1 \bar{E}_2  \,}  \bigg|_{U= U_{\rm OGE}^V + U_{\rm conf}^V  } \,.
\eeq
The tensor term is regarded as small and treated in perturbation theories.

\subsection{Bases and matrix elements}
\label{sec:matrix_elements}

The potential Eq.(\ref{eq:full_expressions_pot}) contains operators of spin, orbital angular momenta and tensor couplings.
The operators ${\bm J}^2$ and $J_z$, made of total angular momentum ${\bm J}={\bm L}+{\bm S}$, commute with all operators in the potential.
Meanwhile, the ${\bm L}^2$ operator is in general not conserved due to the tensor forces that mix, e.g., $L=0$ and $L=2$ states.
Also the total spin is not conserved for mesons including unequal quark masses.
Our strategy is to use bases with definite $(J,J_z,L,S)$ and treat the terms that violate $L$ and $S$ conservations as perturbations.
In this paper we omit such perturbative corrections when we perform fitting to the meson spectra, but 
we summarize some expressions needed to evaluate such matrix elements for the future studies.

\subsubsection{Equal mass}
\label{sec:equal_mass}

The total spin $S$ operator in general does not commute with the $ \sum_{i=1,2} \bar{V}^{(i)}_{ sL } ( {\bm \sigma}_i \cdot \vL ) $ operator.
The exception is the case with equal mass $m_1=m_2$ for which the operator can be expressed as 
\beq
 \sum_{i=1,2} \bar{V}^{(i)}_{ sL } ( {\bm \sigma}_i \cdot \vL ) = \bar{V}_{ sL } ( {\bm \sigma} \cdot \vL )
 \label{eq:av_S_L}
\eeq
with ${\bm \sigma} = {\bm \sigma}_1 + {\bm \sigma}_2$ and $\bar{V}^{(1)}_{ sL }=\bar{V}^{(2)}_{ sL }=\bar{V}_{ sL }$.
Hence, we use the bases ($C^{J,J_z}_{L,L_z; S,S_z }$: Clebsch-Gordan coefficients),
\beq
| J, J_z \ra_{L,S} = \sum_{L_z, S_z } C^{J,J_z}_{L,L_z; S,S_z } | L, L_z \ra \otimes |S,S_z\ra \,. 
\eeq
For mesonic systems, the total spin is $S=0$ or $1$. 
With these bases, the matrix elements of operators (which are diagonal for $(J,J_z, L,S)$) are evaluated.
The spin-spin operator takes the usual form,
\beq
\la {\bm \sigma}_1  \cdot {\bm \sigma}_2  \ra^{J,J_z}_{L,S}
= 2S(S+1) - 3 \,.
\eeq
The spin-orbit operator is evaluated as
\beq
\la {\bm L} \cdot {\bm \sigma} \ra^{J,J_z}_{L,S}
= 2 C_{LS}^{J}  \,,
\label{eq:LS}
\eeq
with
\beq
C_{LS}^{J} \equiv \frac{\, J(J+1) - L(L+1) - S(S+1) \,}{2}  \,.
\eeq
Also, one can compute
\beq
\la ( {\bm L} \cdot {\bm \sigma}_1 ) ( {\bm L} \cdot {\bm \sigma}_2 ) \ra^{J,J_z}_{L,S}
= D_{LS}^J  \,,
\eeq
with 
\beq
D_{LS}^{J} \equiv 2\big(C_{LS}^{J} \big)^2 + C_{LS}^{J} - L(L+1)  \,.
\eeq
%

\subsubsection{Unequal mass}
\label{sec:unequal_mass}

For unequal masses, $S$ is not a good quantum number.
The bases must be a superposition of $S=0$ and $1$ states.
\beq
|J,J_z\ra_L
\equiv \alpha |J,J_z\ra_{L,S=1} + \beta |J,J_z\ra_{L,S=0} \,.
\eeq
For a given $L$, the possible $J$'s are $L+1, L$, and $|L-1|$.
The cases $J=L+1$ and $|L-1|$ are possible only when $S=1$, so $\alpha=1$ and $\beta=0$.
The matrix elements for ${\bm L}\cdot {\bm \sigma}_{1,2}$ are 
\beq
&&
\la L+1, J_z | ( \vL \cdot {\bm \sigma}_{1,2} ) |  L+1, J_z \ra_{L,S=1} = L \,,
\nonumber \\
&&
\la L-1, J_z | ( \vL \cdot {\bm \sigma}_{1,2} ) | L-1, J_z \ra_{L,S=1} = -(L+1) \,,
\eeq
where the expression for $L-1$ is valid for $L\ge 1$.

For $J=L$, we determine $\alpha$ and $\beta$ for a given $(J,L)$ by diagonalizing the hamiltonian with the following mixing terms
(the upper are $(S,S')=(1,1)$ and $(1,0)$ components, the lower $(0,1)$ and $(0,0)$ components),
\beq
&&  \bigg\la \sum_{i=1,2} \bar{V}^{(i)}_{ sL } ( {\bm \sigma}_i \cdot \vL ) \bigg\ra^{J=L, L}_{S,S'}
\nonumber \\
&&
= \left[
\begin{matrix}
 - \bar{V}^{(1+2)}_{ sL }  & - \sqrt{L(L+1)\,} \, \bar{V}^{(1-2)}_{ sL } \, \\
\, - \sqrt{L(L+1)\,} \, \bar{V}^{(1-2)}_{ sL }  \, & 0
\end{matrix}
\right]^{J=L,L}
\nonumber \\
\eeq
where $\bar{V}^{(1\pm 2)}_{ sL } =  \bar{V}^{(1)}_{ sL } \pm  \bar{V}^{(2)}_{ sL }$. 
Here the $L=0$ case is exceptional because of complete decoupling of $S=1$ and $0$ states.

\begin{table}[htb]
\caption{
Meson spectra for the $L=0$ states from light to charm quark sectors. 
The experimental meson mass $M_{\rm exp}$, our result $M_{\rm th}$, the average momenta which satisfy the consistency, $\bar{\vP}^2 \simeq \la \vP^2 \ra$,
are given in GeV units, while the root-mean-square radii $\sqrt{\la \vecr^2 \ra \,}$ are given in fm unit.
Our choices for $f_S$ and $\alpha_s$ are also displayed.
(* for unestablished states.)
}
\label{tab:spectra_L0}       
\begin{tabular}{c c c c c c c c c c c c c c c}
\hline\noalign{\smallskip}
 & $n_r$$^{2S+1} L_J$ & $ M_{\rm exp} $ & $ M_{\rm cal} $  &~ $\bar{\vP}^2$ ~& $\sqrt{ \la \vecr^2\ra}$  &~ $f_S$ &~ $\alpha_s$
\\
\hline\noalign{\smallskip}
$\pi$ &~$1^1 S_0$          &~ 0.14    ~&~ 0.16 ~&~~0.47~ &~0.50 &~ 0.70  &~ 0.80 \\
         &~$2^1 S_0$          &~ 1.30    ~&~ 1.28 ~&~~0.43~ &~0.98 &~ ~ &~  \\
         &~$3^1 S_0$          &~  1.81   ~&~ 1.82 ~&~~0.55~ &~1.38 &~ ~ &~  \\
         &~$4^1 S_0$          &~  2.07**   ~&~ 2.22 ~&~~0.67~ &~1.66 &~ ~ &~  \\         
         \hline\noalign{\smallskip}           
$\rho$ &~$1^3 S_1$          &~ 0.78     ~&~ 0.76 ~&~~0.21~ &~0.66 &~ 0.74  &~ 0.80  \\
           &~$2^3 S_1$          &~ 1.47      ~&~ 1.44 ~&~~0.35~ &~1.17 & ~ \\
           &~$3^3 S_1$          &~  1.91*      ~&~ 1.87 ~&~~0.48~ &~1.55 &~ ~ \\ 
            &~$4^1 S_1$          &~  2.27**   ~&~ 2.22 ~&~~0.61~ &~1.83 &~ ~ &~  \\     
           \hline\noalign{\smallskip}
                    \hline\noalign{\smallskip} 
$K$     &~$1^1 S_0$          &~ 0.49   ~&~ 0.49 ~&~~0.42~ &~0.49 &~ 0.72 &~ 0.77 \\
           &~$2^1 S_0$          &~ 1.46*  ~&~ 1.46 ~&~~0.45~ &~0.98 & ~ \\
                      \hline\noalign{\smallskip}
$K^*$     &~$1^3 S_1$      &~ 0.89   ~&~ 0.91 ~&~~0.24~ &~0.63 &~ 0.75 &~ 0.77 \\
           &~$2^3 S_1$          &~ 1.41      ~&~ 1.54 ~&~~0.39~ &~1.10 &~ ~ \\
           \hline\noalign{\smallskip}
                    \hline\noalign{\smallskip} 
$\eta_{s}$\footnote{We assume $ |\eta \ra = \frac{ |u \bar{u} \rangle + |d \bar{d} \rangle - 2|s \bar{s} \rangle }{\sqrt{6} } =  \frac{ \sqrt{2}  | \pi \rangle - 2| \eta_s \rangle }{\sqrt{6} } $ to derive the mass relation $m_{\eta} = \frac{\, m_\pi + 2 m_{\eta_s}}{3} $. }     
                      &~$1^1 S_0$    &~ 0.74   ~&~ 0.71 ~&~~0.45~ &~0.47 &~ 0.73 &~ 0.75 \\
                       &~$2^1 S_0$    &~ 1.48     ~&~ 1.66 ~&~~0.46~ &~0.99 &         &   \\
                        &~$3^1 S_0$    &~ 2.10     ~&~ 2.12 ~&~~0.59~ &~1.37 &         &   \\
                      \hline\noalign{\smallskip}
$\phi$     &~$1^3 S_1$      &~ 1.02   ~&~ 1.03 ~&~~0.30~ &~0.57 &~ 0.76 &~ 0.75\\
               &~$2^3 S_1$      &~ 1.68  ~&~ 1.71 ~&~~0.43~ &~1.06 &~ ~ & \\
               &~$3^1 S_1$    &~ 2.18     ~&~ 2.12 ~&~~0.56~ &~1.44 &         &   \\
           \hline\noalign{\smallskip}
                    \hline\noalign{\smallskip} 
$D$     &~$1^1 S_0$          &~ 1.87   ~&~ 1.87 ~&~~0.39~ &~0.50 &~ 0.73 &~ 0.58 \\
           &~$2^1 S_0$          &~ ---    ~&~ 2.50 ~&~~0.47~ &~0.99 & ~ & \\
                      \hline\noalign{\smallskip}
$D^*$     &~$1^3 S_1$      &~ 2.01   ~&~ 2.00 ~&~~0.29~ &~0.56 &~ 0.76 &~ 0.58 \\
           &~$2^3 S_1$          &~ 2.64*  ~&~ 2.53 ~&~~0.44~ &~1.04 & & \\
           \hline\noalign{\smallskip}
                    \hline\noalign{\smallskip} 
$D_s$     &~$1^1 S_0$      &~ 1.97   ~&~ 1.96 ~&~~0.52~ &~0.44 &~ 0.73 &~ 0.57 \\
           &~$2^1 S_0$          &~ ---      ~&~ 2.66 ~&~~0.56~ &~0.91 &~ ~ & \\
                      \hline\noalign{\smallskip}
$D_s^*$     &~$1^3 S_1$   &~ 2.11   ~&~ 2.12 ~&~~0.37~ &~0.50 &~ 0.76 &~ 0.57 \\
           &~$2^3 S_1$          &~ 2.73   ~&~ 2.70 ~&~~0.51~ &~0.96 &~ & \\
         \hline\noalign{\smallskip}          
                             \hline\noalign{\smallskip} 
$\eta_c$     &~$1^1 S_0$   &~ 2.98   ~&~ 2.98 ~&~~0.79~   &~0.35 &~ 0.73 &~ 0.42 \\
           &~$2^1 S_0$          &~ 3.64      ~&~ 3.62 ~&~~0.81~ &~0.78 &~ ~ & \\
                      \hline\noalign{\smallskip}
$J/\psi$     &~$1^3 S_1$      &~ 3.10   ~&~ 3.09 ~&~~0.56~ &~0.40 &~ 0.77 &~ 0.42 \\
 $\psi$          &~$2^3 S_1$   &~ 3.69   ~&~ 3.66 ~&~~0.71~ &~0.82 & & \\
	\hline\noalign{\smallskip}
	\hline\noalign{\smallskip}
\end{tabular}
\vspace*{0.1cm}  
\end{table}

\begin{table}[htb]
\caption{
The average quark energy vs composition of the potential energies for mesons with equal quark masses, $m_1=m_2$. The unit is GeV.
We use the following conventions for the evaluation of potential energies: 
$\la O_C^{(0,2)} \ra = \la \bar{V}_C^{(0,2)} \ra$ and $\la O_{ss} \ra = \la \bar{V}_{ss} ( {\bm \sigma}_1 \cdot {\bm \sigma}_2 ) \ra$.
}
\label{tab:composition_L0}       
\begin{tabular}{c c c c c c c c c c c c c c c}
\hline\noalign{\smallskip}
 & $ M_{\rm cal} $  &~ $\bar{\vP}^2 $~ &~ $\bar{E}_{1,2}$ ~&~ $\la O_C^{(0)} \ra$  &~ $\la O_C^{(2)} \ra$ &~ $\la O_{ss} \ra $
\\
\hline\noalign{\smallskip}
$\pi$               &~ 0.16 ~&~~0.47~ &~$0.75$       &~$-1.01$    &~$0.10$   &~$-0.34$ \\
                       &~ 1.28 ~&~~0.43~ &~$0.72$       &~$-0.07$    &~$0.03$   &~$-0.12$ \\
                   \hline\noalign{\smallskip}           
$\rho$             &~ 0.76 ~&~~0.21~ &~$0.54$   &~$-0.43$  &~$0.05$  &~$0.05$   \\
                       &~ 1.44 ~&~~0.35~ &~$0.66$   &~$0.06$   &~$0.04$  &~$0.04$   \\
            \hline\noalign{\smallskip}
                    \hline\noalign{\smallskip} 
$\eta_{s}$     &~ 0.71 ~&~~0.45~ &~$0.83$    &~$-0.82$ &~$0.07$  &~$-0.17$ \\
                       &~ 1.66 ~&~~0.46~ &~$0.83$    &~$0.03$   &~$0.03$  &~$-0.05$ \\
                      \hline\noalign{\smallskip}
$\phi$             &~ 1.03 ~&~~0.30~ &~$0.73$    &~$-0.49$    &~$0.05$    &~$0.04$ \\
                       &~ 1.71 ~&~~0.43~ &~$0.82$    &~$0.05$     &~$0.03$    &~$0.02$ \\
           \hline\noalign{\smallskip}
                    \hline\noalign{\smallskip} 
$\eta_c$    &~ 2.98 ~&~~0.79~   &~$1.70$     &~$ -0.37$     &~$0.02$    &~$-0.07$ \\
                  &~ 3.62 ~&~~0.81~   &~$1.71$     &~$0.26$       &~$0.01$    &~$-0.04$ \\
                      \hline\noalign{\smallskip}
$J/\psi$     &~ 3.09 ~&~~0.56~ &~$1.63$      &~$-0.20$       &~$0.01$    &~$0.01$ \\
 $\psi$       &~ 3.66 ~&~~0.71~ &~$1.67$      &~$0.31$        &~$0.01$    &~$0.01$ \\
	\hline\noalign{\smallskip}
	\hline\noalign{\smallskip}
\end{tabular}
\vspace*{0.1cm}  
\end{table}

\section{Meson spectra }
\label{sec:meson_spectra}

\begin{figure}[bt]
\vspace{-0.cm}
\begin{center}	
	\includegraphics[width=8.5cm]{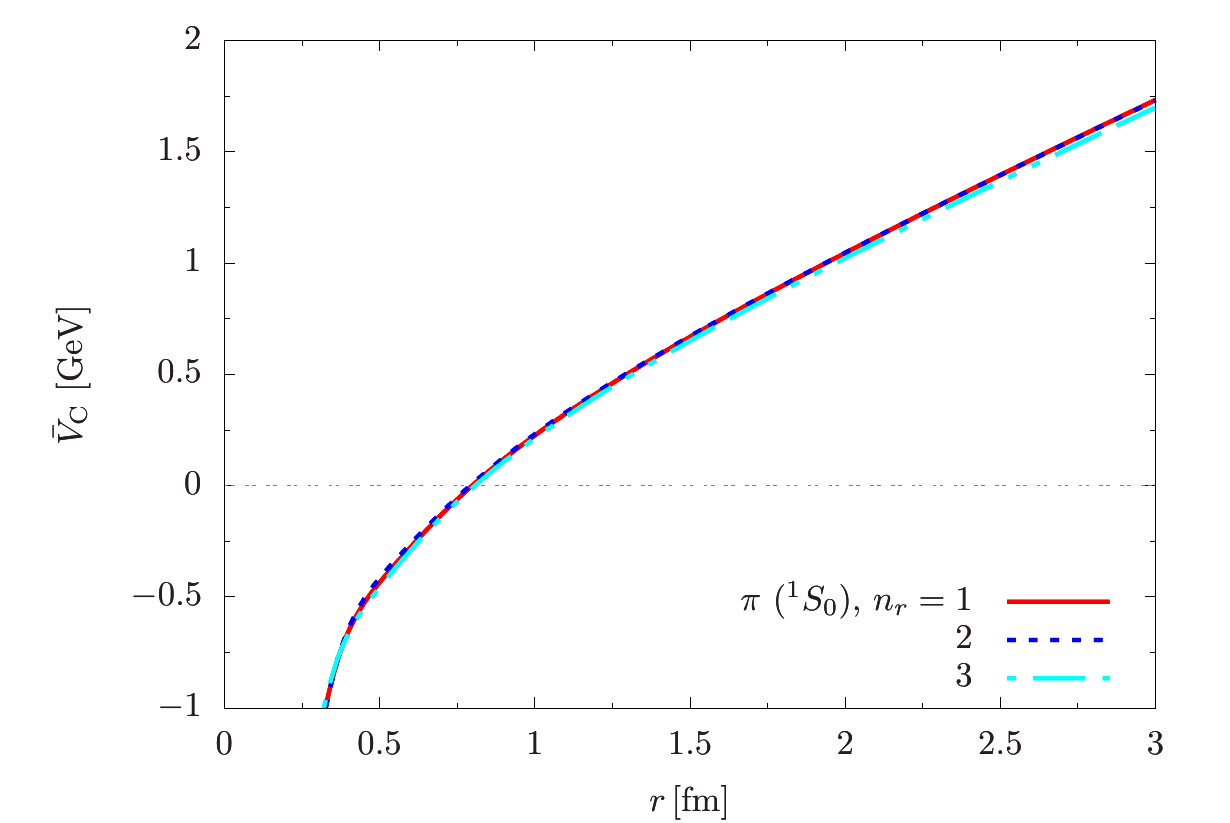}
	\includegraphics[width=8.5cm]{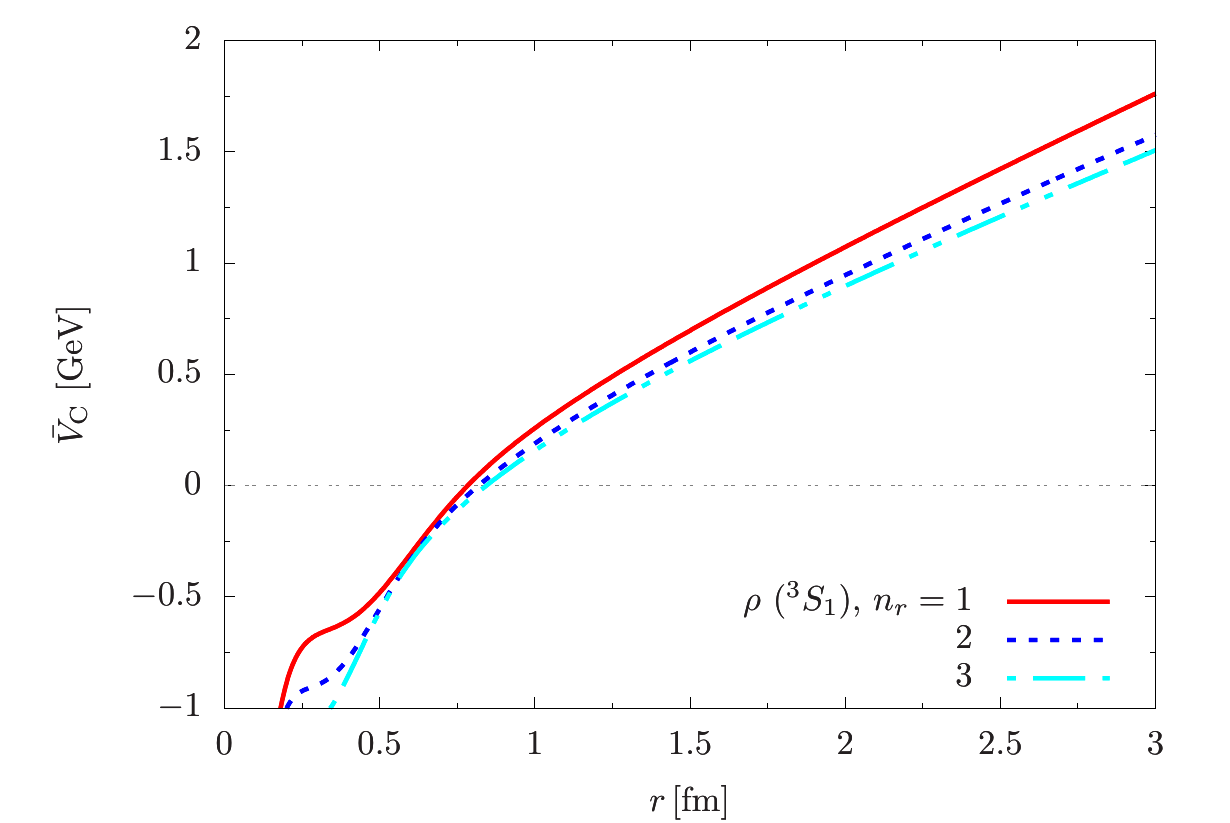}

		\vspace{-0.2cm}
\caption{  The potential $\bar{V}_C$ for $\pi\, (^1S_0)$ (upper panel) and $\rho\, (^3S_1)$ (lower panel) states with radial excitations up to $n_r=3$. The potential depends on the average momentum $\bar{\vP}^2$ in a given meson. Small bumps in the $\rho$ channel originate from the spin-spin repulsions at short distance.
		 }
		 		\vspace{-0.4cm}
		 \label{fig:V_central_ud}
\end{center}		
		
\end{figure}
\begin{figure}[bt]
\vspace{-0.cm}
\begin{center}	
	\includegraphics[width=8.5cm]{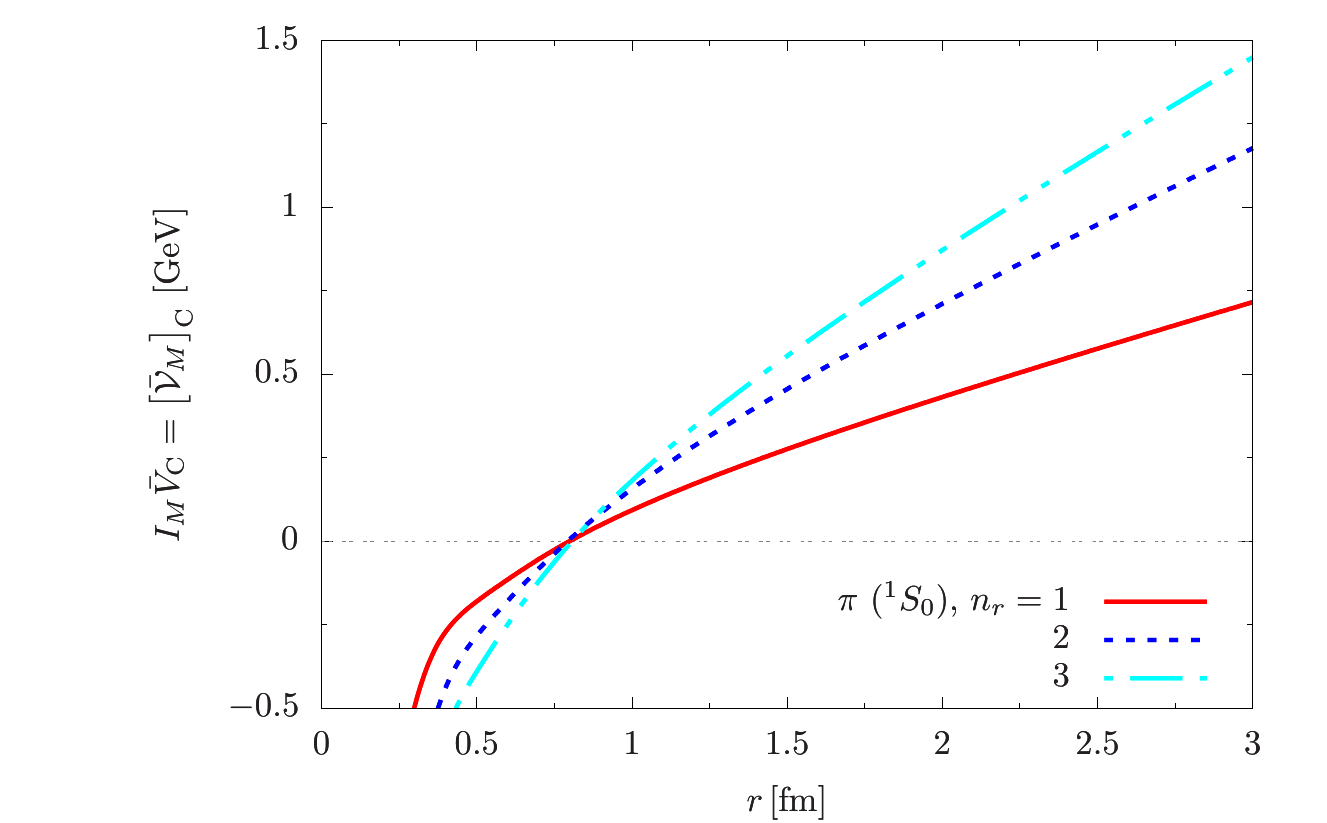}
	\includegraphics[width=8.5cm]{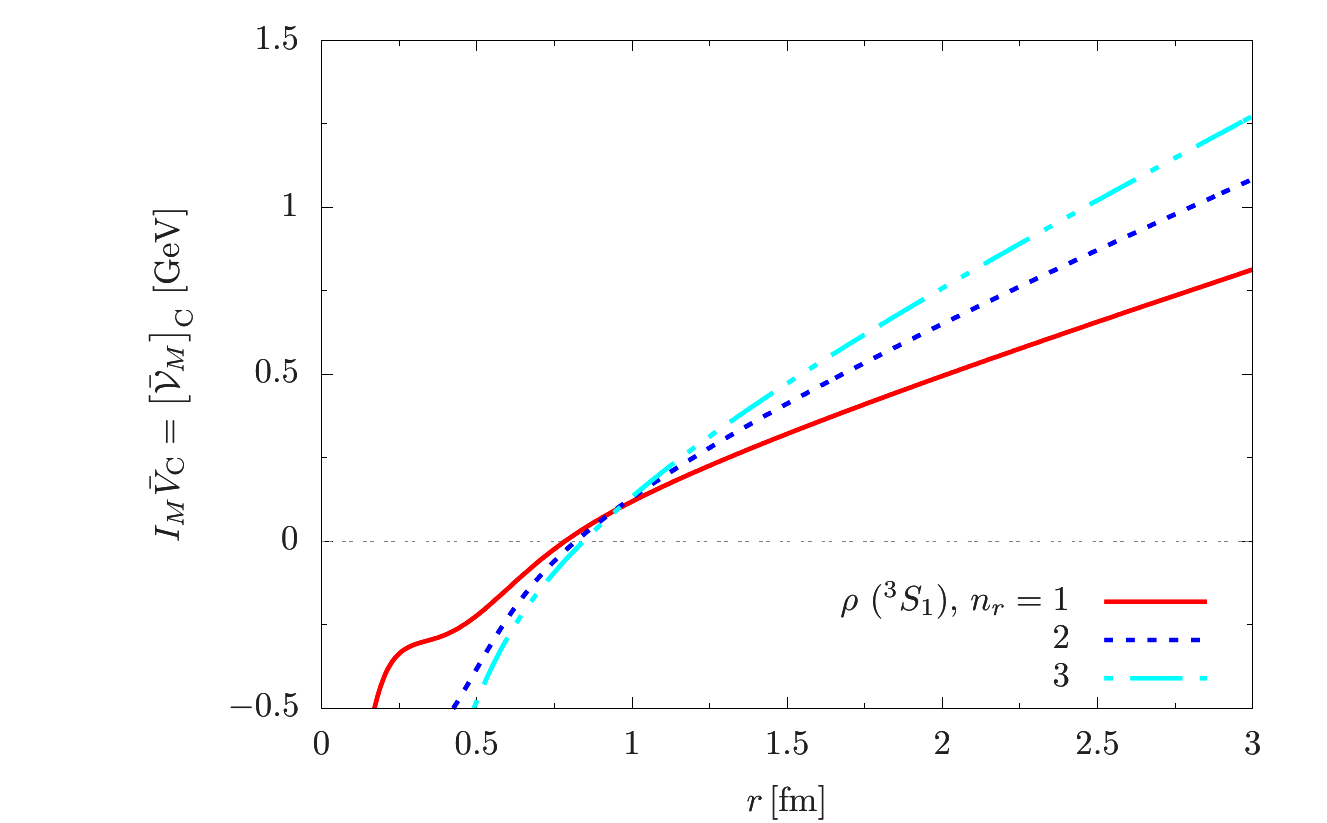}

		\vspace{-0.2cm}
\caption{  The effective potential $[\calV_M]_{C} = I_M \bar{V}_C$ for $\pi\, (^1S_0)$ (upper panel) and $\rho\, (^3S_1)$ (lower panel) states with radial excitations up to $n_r=3$. 
The effective potential includes the relativistic kinematic factor $I_M$ in Eq.(\ref{eq:I_M}) which depend on the meson mass and $\bar{E}_{1,2}$.
In the nonrelativistic limit for equal masses, the factor $I_M$ is a constant, $I_M = m_1=m_2$. 
		 }
		 		\vspace{-0.4cm}
		 \label{fig:calV_central_ud}
\end{center}		
		
\end{figure}

In this section we fit our model calculations to the experimental meson spectra.
We focus on ordinary mesons which are regarded as $q \bar{q}$, omitting extraordinary mesons such as $\eta'$, 
which is substantially affected by the topological fluctuations \cite{tHooft:1976rip,tHooft:1976snw,Witten:1979vv},
or scalar mesons, $\sigma, \kappa, f_0(980), a_0(980)$, 
which are often discussed as tetraquark candidates including intermediate states with the $q\bar{q}$ annihilations or mesonic molecule states \cite{Jaffe:1976ig,Isgur:1976qg,Pelaez:2015qba}.

There are several states for which assignments of quantum numbers are not obvious.
For mesons made of light and strange quarks, 
we follow the identification made by Ebert et al. where the list of mesons is summarized in Table I in Ref.\cite{Ebert:2009ub}.
For light-heavy mesons, we refer to Table 1 in Ref.\cite{Ebert:2009ua}.
If the assigned states are not well established in the particle data group (PDG) \cite{ParticleDataGroup:2020ssz}, 
we attach * to the experimental mass $M_{\rm exp}$.
We attach ** for states in ``further states'' Section in the PDG.

As we have seen in the previous sections, our model contains many parameters which cannot be directly derived from our model. The list is
\beq
(m_{ud},\, m_s,\, m_c),\, f_S,\, \alpha_s,\, 
C_{\calK_i},\,   C_{\calL_i},\, C_{\calJ_1 \calJ_2} \,.
\eeq
Moreover these parameters may depend on dynamics and thus may change for a given meson.
We try to keep these parameters common for different channels as much as possible. 
But we found it necessary to make some parameters non-universal. 
Below we examine when we need to compromise with non-universal treatments.

We take the constituent quark masses common for whole mesons,
\beq
(m_{ud},\, m_s,\, m_c) = (0.30,\, 0.48,\, 1.45)\, {\rm GeV}\,,
\eeq
which are typical in constituent quark models. Some variations are possible and the effects can be largely absorbed by retuning the values of $\alpha_s$.

We found that the value of $f_S$ cannot be taken common for whole spectra and should vary from $\simeq 0.70$ to $\simeq 0.83$. 
Light flavor mesons favor smaller values.
The trend $f_s > 0.5$ means that the scalar confinement vertex is larger than the vector one. 
In this respect our confining part is closer to Godfrey et al. with $f_S=1$ than to Ebert et al. with $f_S=-1$.

The $\alpha_s$ values are sensitive to the flavors of mesons which are related to the quark energies.
For our choice of the quark masses, $\alpha_s$ ranges from $\simeq 0.8$ for light quarks to $\simeq 0.3$ for charm quarks. 
For mesons with unequal quark masses $m_f \neq m_{f'}$, we found that the following estimate works reasonably well for the effective coupling $\alpha_s^{ff'}$,
\beq
\alpha_s^{ff'} \simeq \sqrt{\alpha_s^f \alpha_s^{f'} }\,,
\eeq
where $\alpha_s^f$ is the value used for a meson made of $m_1=m_2=m_f$.
The details depend on the channels, and we allow variation of $\lesssim 10$\% in our fitting procedures.

For $C$'s appearing in the form factors, it should be $O(1)$ by construction.
We found that $C_{\calJ_1 \calJ_2}$ and $C_{\calL_i}$, which appear in spin-spin forces and $LS$ respectively, need some arrangements.
Meanwhile, the spectra are not very sensitive to $C_{\calK_i}$  as far as we take the values close to $\simeq 1$.
We choose
\beq
C_{\calL_i} = 0.5\,,~~~C_{\calK_i} = 1\,,~~~ C_{\calJ_1 \calJ_2} = 1.0-1.5 \,.
\eeq
We found that the values of $C_{\calJ_1 \calJ_2}$ should deviate from $1$ for mesons including light quarks, i.e., $ud$, $us$, and $ss$ mesons for good fits.

In our fitting procedures we omit tensor forces (which mix different $L$) and $({\bm s}_1-\bm{s}_2) \cdot \vL$ type LS forces (which mix $S=0$ and 1 states).
While these terms make our fitting procedures more nonlinear and involved, we expect that their impacts are not very significant 
at the quality of fit we are aiming at in this paper.

\subsection{$L=0$ states}
\label{sec:fit_L=0}

\begin{figure*}[htb]
\vspace{-0.cm}
\begin{center}	
\includegraphics[width=18.6cm]{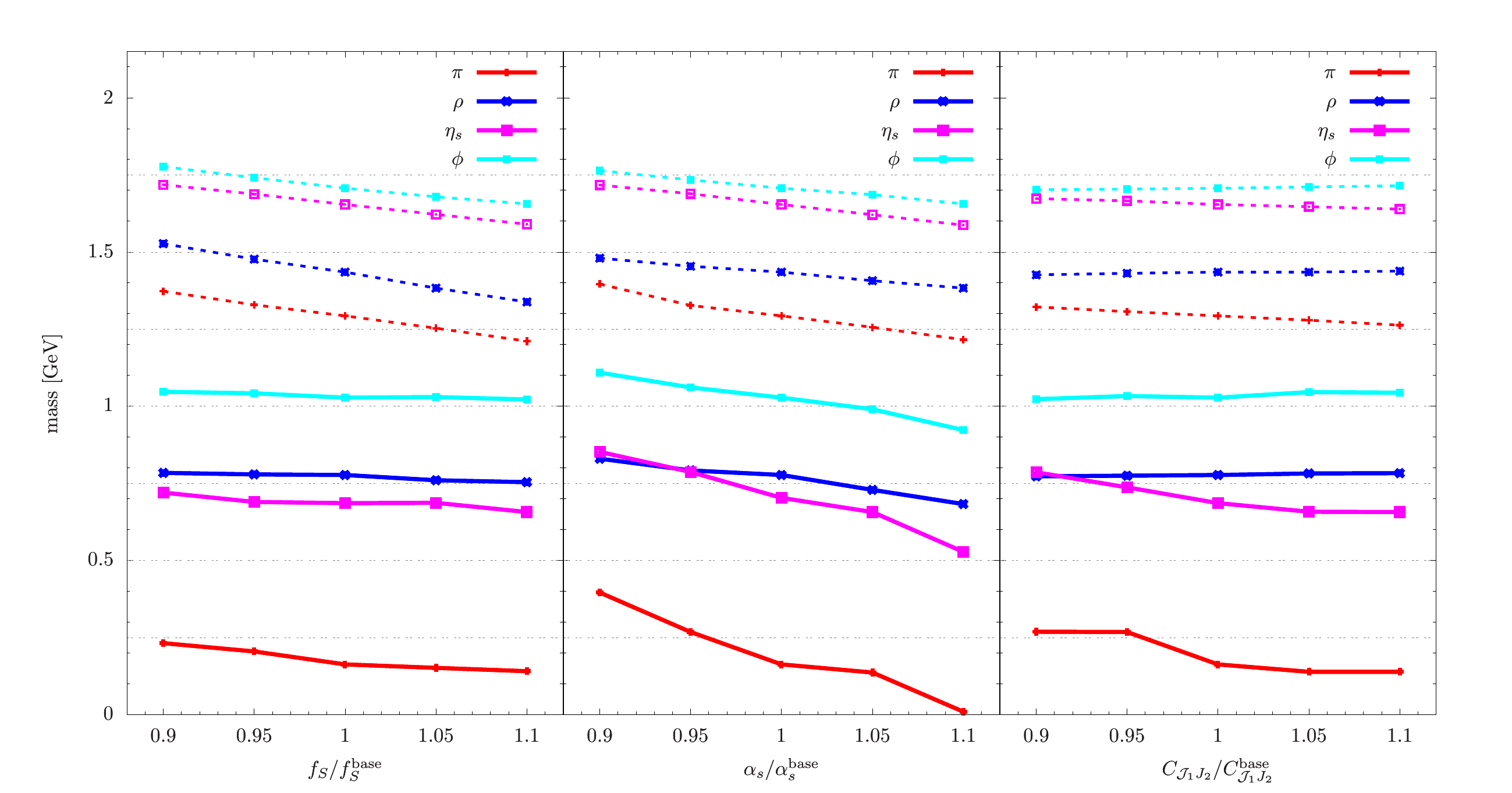}
\caption{  The $f_S, \alpha_s, C_{\calJ_1 \calJ_2}$ dependence of the masses of $\pi,\rho, \eta_{s}, \phi$ for the 0th (solid lines) and the 1st radial excitations (dashed lines). 
The superscript ``base'' refers to the baseline values which we chose for Table.\ref{tab:spectra_L0}.
Each parameter is varied up to $\pm 10$\% from the baseline values.
}
 \label{fig:para_vary}
\end{center}		
\vspace{+0.cm}		
\end{figure*}

The $L=0$ states depend on the central and spin-spin potentials,
\beq
\bar{V}_{L=0} 
&=&
 \bar{V}_{ C }^{(0)} + \bar{V}_{ C }^{(2)} + \bar{V}_{ ss}  ( {\bm \sigma}_1 \cdot {\bm \sigma}_2 ) \,.
\label{eq:full_L=0_meson}
\eeq
In order to achieve good fits, we take $C_{\calJ_1\calJ_2} = 1.5$ for the $ud$-mesons,  $C_{\calJ_1\calJ_2} = 1.3$ for the $us$-mesons, and for the others we set $C_{\calJ_1 \calJ_2}=1$.
The parameter $C_{\calL_i}$ does not show up in the $L=0$ states and are left unfixed.

In Table.\ref{tab:spectra_L0} we display 
the experimental meson mass $M_{\rm exp}$, our result $M_{\rm th}$, the average momenta which satisfy the consistency, $\bar{\vP}^2 \simeq \la \vP^2 \ra$,
and the root-mean-square radii $\sqrt{\la \vecr^2 \ra \,}$.
In Table.\ref{tab:composition_L0} we show the composition of the potential energies for mesons with equal mass quarks,
together with the average quark energy $\bar{E}_{1,2}$.
The calculated masses fit the experimental ones very well from the light to charm quark sectors,
although this is probably not very surprising because we have many parameters. 
What is important here is to examine the impact of each parameter and its trend.

Shown in Fig.\ref{fig:V_central_ud} are the potentials $\bar{V}_C$ which depend on meson masses and $\bar{\vP}^2$. 
The cases for $\pi\, (^1S_0)$ (upper panel) and $\rho\, (^3S_1)$ (lower panel) states with radial excitations up to $n_r=3$ are displayed. 
Small bumps in the $\rho$ channel originate from the spin-spin repulsions at short distance.
For our choice of $f_S\simeq 0.70$-0.75, it turns out that the potentials do not show large sensitivity to the values of $\bar{\vP}^2$.  
In Fig.\ref{fig:calV_central_ud}, we show the potentials $[\bar{\calV}_M]_C$ where the relativistic kinematic factors $I_M$'s in Eq.(\ref{eq:I_M}) are multiplied to $\bar{V}_C$. 
For equal quark masses $m_1=m_2$, the factor $I_M = (M+ \bar{E}_1+\bar{E}_2)/4$ depends on the meson mass and the quark energy.
It becomes larger for a heavier meson, unlike the non-relativistic case where $I_M=m_{1,2}$ is a constant (see, Eq.(\ref{eq:Vhh})).

Given our fits, we found general trends:
(i) To reproduce the masses of excited states, we had to tune $f_S$ for given channels.
The trends we found are that $f_S$ should be taken smaller for the spin-singlet states than the triplet case, and $f_S$ gently grows toward the charm quark sector.
(ii) For a given flavor combination, it turns out that we can fit the spectra using the common $\alpha_s$, although it is not necessary from the physics point of view.
The value of $\alpha_s$ decreases toward the heavy quark sector, as it should.
(iii) The momenta are in general sizable compared to the quark masses, even for the charm quark case.
The resultant large average quark energies are largely cancelled with the Coulomb attraction energies $\bar{V}_C^{(0)}$.
Due to this structure, the choice of $\alpha_s$ is crucial.
(iv) The second order central potential $\bar{V}_C^{(2)}$ is much smaller than the leading order.
This corrections contain the uncertainties associated with the choice of $C_{\calK_i}$, but the details are not very important and thus we simply take $C_{\calK_i}=1$.
(v) The spin-spin potential $\bar{V}_{ss}$ is sizable for the $n_r=0$, spin-singlet states as in conventional non-relativistic models.
Meanwhile, the spin-spin repulsion in the triplet channel is not as large as in the non-relativistic models.

While the above qualitative trends are commonly found for various parameter sets,
the quantitative details of the mass spectra are rather sensitive to the parameters.
To examine the importance of fine-tuning for parameters $f_S, \alpha_s$, and $C_{\calJ_1 \calJ_2}$, 
we perform linear analyses in which we change one of these parameters by $\sim 10$\% while fix all the other parameters.
In Fig.\ref{fig:para_vary} we show the results for $\pi, \rho, \eta_{s}, \phi$ including the 0th and 1st radial excitations.
During this analysis we always arrange $\bar{\vP}^2 \simeq \la \vP^2 \ra$ in good accuracy within our grids, $\Delta \bar{\vP}^2 =0.01\,{\rm GeV}^2$.

The correlations among these three parameters and the spectra are relatively simple to understand.
The 10\% variation of $f_S$ introduces the marginal changes in the 0th radial excitations while induces the mass changes of $\sim 200$ MeV for the 1st radial excitations.
Mesons with heavier flavors have more compact structures and thus are less affected by the details of confining potentials.

The 10\% variation of $\alpha_s$ has very large effects on the spin-singlet states. For $\pi$ and $\eta_s$, the mass variation is $\sim 300$-$400$ MeV.
For the spin triplet states, $\rho$ and $\phi$, the impacts are smaller but still are $\sim 100$-$200$ MeV. 
The difference between the singlet and triplet states is the spin-spin correlations which make mesons more compact for the spin-singlet states;
$\pi$ and $\eta_s$ have more chances to reduce the masses by forming the compact wavefunctions.
For heavier flavors or higher radial excitations, the effects of $\alpha_s$ become weaker.

The 10\% variation of $C_{\calJ_1 \calJ_2}$ has large impacts on the spin singlet states, while the spin triplet states are not much affected.

After observing these correlations among the parameters and the spectra, our fitting strategy has been fixed as follows.
We first explored the reasonable range of $\alpha_s$ to fit the spin triplet states in the 0th radial excitations
which are relatively insensitive to $f_S$ and $C_{\calJ_1 \calJ_2}$.
Then, we tuned $C_{\calJ_1 \calJ_2}$ to fit the spin singlet states.
In the third step we chose $f_S$ to fit the radial excitations.
The rest requires the fine tuning of all these parameters slightly.
The last step is to examine whether the chosen parameters show qualitatively acceptable trends.
We believe that the parameters shown in Table.\ref{tab:spectra_L0} are reasonable.
For instance, the range of $\alpha_s$ used in the fit is consistent with Eq.(\ref{eq:alpha_s_para}).

\subsection{$L\ge1$ states}
\label{sec:fit_L=1}

\begin{table}[ptb]
\caption{
The meson spectra for the $L=1$ states with the conventions same as Table.\ref{tab:spectra_L0}. 
From top to bottom, $ud$, $us$, $ss$, $uc$, $us$, and $cc$ mesons.
The scalar mesons $^3 P_0$ are omitted.
}
\label{tab:spectra_L1}       
\begin{tabular}{c c c c c c c c c c c c c c c}
\hline\noalign{\smallskip}
 & $n_r$$^{2S+1} L_J$ & $ M_{\rm exp} $ & $ M_{\rm cal} $  &~ $\bar{\vP}^2$ ~& $\sqrt{ \la \vecr^2\ra}$  &~ $f_S$ &~ $\alpha_s$
\\
\hline\noalign{\smallskip}
$b_1$ &~$1^1 P_1$          &~ 1.23    ~&~ 1.27 ~&~~0.25~ &~0.99 &~ 0.77  &~ 0.80 \\
           &~$2^1 P_1$          &~ ---        ~&~ 1.70 ~&~~0.37~ &~1.46 &~ ~ &~  \\
            &~$3^1 P_1$          &~ 1.96**   ~&~ 2.06 ~&~~0.48~ &~1.82 &~ ~ &~  \\
         \hline\noalign{\smallskip}           
$a_1$ &~$1^3 P_1$          &~ 1.23     ~&~ 1.19 ~&~~0.28~ &~0.94 &~   &~   \\
           &~$2^3 P_1$          &~ 1.65      ~&~ 1.68 ~&~~0.41~ &~1.44 & ~ \\
           &~$3^3 P_1$          &~ 2.10**      ~&~ 2.03 ~&~~0.48~ &~1.81 & ~ \\
         \hline\noalign{\smallskip}                    
$a_2$ &~$1^3 P_2$          &~ 1.32     ~&~ 1.33 ~&~~0.23~ &~1.03 &~   &~   \\
           &~$2^3 P_2$          &~ 1.73     ~&~ 1.74 ~&~~0.36~ &~1.47 & ~ \\
           &~$3^3 P_2$          &~ 2.05**     ~&~ 2.07 ~&~~0.47~ &~1.84 & ~ \\
	\hline\noalign{\smallskip}
	\hline\noalign{\smallskip}
$K_1$ &~$1 P_1$          &~ 1.40    ~&~ 1.40 ~&~~0.28~ &~0.95 &~ 0.80  &~ 0.79 \\
           &~$2 P_1$          &~ ---        ~&~ 1.80 ~&~~0.44~ &~1.35 &~ ~ &~  \\
         \hline\noalign{\smallskip}            
$K_1$ &~$1 P_1$          &~ 1.27    ~&~ 1.31 ~&~~0.30~ &~0.90 &~   &~  \\
           &~$2 P_1$          &~ 1.65*    ~&~ 1.77 ~&~~0.42~ &~1.35 &~ ~ &~  \\
         \hline\noalign{\smallskip}            
$K_2^*$ &~$1^3 P_2$       &~ 1.42    ~&~ 1.47 ~&~~0.25~ &~1.01 &~   &~  \\
           &~$2^3 P_2$          &~ 1.98*   ~&~ 1.85 ~&~~0.41~ &~1.39 &~ ~ &~  \\
         \hline\noalign{\smallskip}     
         \hline\noalign{\smallskip}
$h_1$ &~$1^1 P_1$          &~ 1.39    ~&~ 1.55 ~&~~0.33~ &~0.80 &~ 0.80  &~ 0.78 \\
           &~$2^1 P_1$          &~ 1.60*   ~&~ 1.70 ~&~~0.42~ &~1.37 &~ ~ &~  \\
         \hline\noalign{\smallskip}
$f_1$ &~$1^3 P_1$          &~ 1.43    ~&~ 1.44 ~&~~0.42~ &~0.80 &~   &~  \\
           &~$2^3 P_1$          &~ 1.97*  ~&~ 1.88 ~&~~0.54~ &~1.24 &~ ~ &~  \\
        \hline\noalign{\smallskip}
$f'_2$ &~$1^3 P_2$          &~ 1.53   ~&~ 1.62 ~&~~0.28~ &~0.94 &~   &~  \\
           &~$2^3 P_2$          &~ 2.01        ~&~ 2.03 ~&~~0.42~ &~1.36 &~ ~ &~  \\
         \hline\noalign{\smallskip}     
         \hline\noalign{\smallskip}
$D_1$ &~$1 P_1$           &~ 2.42    ~&~ 2.40 ~&~~0.34~ &~0.86 &~ 0.80  &~ 0.55 \\
           &~$2 P_1$           &~ ---        ~&~ 2.72 ~&~~0.49~ &~1.27 &~ ~ &~  \\
         \hline\noalign{\smallskip}
$D_1$ &~$1 P_1$           &~ 2.43    ~&~ 2.36 ~&~~0.38~ &~0.81 &~   &~  \\
           &~$2 P_1$           &~ ---        ~&~ 2.72 ~&~~0.51~ &~1.24 &~ ~ &~  \\   
        \hline\noalign{\smallskip}
$D_2^*$ &~$1^3 P_2$     &~ 2.46    ~&~ 2.44 ~&~~0.30~ &~0.90 &~   &~  \\
               &~$2^3 P_2$     &~ ---        ~&~ 2.79 ~&~~0.45~ &~1.31 &~ ~ &~  \\  
         \hline\noalign{\smallskip}     
         \hline\noalign{\smallskip}
$D_{s1}$ &~$1 P_1$       &~ 2.54    ~&~ 2.57 ~&~~0.41~ &~0.78 &~ 0.80  &~ 0.55 \\
           &~$2 P_1$            &~ ---        ~&~ 2.95 ~&~~0.57~ &~1.18 &~ ~ &~  \\
                    \hline\noalign{\smallskip}
$D_{s1}$ &~$1 P_1$       &~ 2.46    ~&~ 2.52 ~&~~0.46~ &~0.75 &~   &~ \\
           &~$2 P_1$            &~ ---        ~&~ 2.91 ~&~~0.60~ &~1.15 &~ ~ &~  \\
                    \hline\noalign{\smallskip}
$D_{s2}$ &~$1^3 P_2$       &~ 2.57    ~&~ 2.60 ~&~~0.37~ &~0.81 &~   &~  \\
           &~$2^3 P_2$            &~ ---       ~&~ 2.98 ~&~~0.53~ &~1.21 &~ ~ &~  \\           
         \hline\noalign{\smallskip}     
         \hline\noalign{\smallskip}                
$h_{c}$       &~$1^1 P_1$       &~ 3.53    ~&~ 3.53 ~&~~0.59~ &~0.65 &~ 0.80  &~ 0.40 \\
                   &~$2^1 P_1$         &~ ---        ~&~ 3.92 ~&~~0.80~ &~0.99 &~ ~ &~  \\
                    \hline\noalign{\smallskip}         
$\chi_{c1}$ &~$1^3 P_1$       &~ 3.51    ~&~ 3.51 ~&~~0.63~ &~0.63 &~   &~  \\
                   &~$2^3 P_1$         &~ 3.87        ~&~ 3.90 ~&~~0.83~ &~0.98 &~ ~ &~  \\
                  \hline\noalign{\smallskip}  
$\chi_{c2}$ &~$1^3 P_2$       &~ 3.56    ~&~ 3.54 ~&~~0.58~ &~0.65 &~   &~  \\
                   &~$2^3 P_2$         &~ 3.93        ~&~ 3.93 ~&~~0.77~ &~1.00 &~ ~ &~  \\ 
         \hline\noalign{\smallskip}     
         \hline\noalign{\smallskip}    
                                     
\end{tabular}
\vspace*{0.1cm}  
\end{table}

\begin{table}[ptb]
\caption{
The meson spectra for the $L=2$ states with the conventions same as Table.\ref{tab:spectra_L0}. 
From top to bottom, $ud$, $us$, $ss$  mesons. Most of mesons with charm quarks are not experimentally confirmed, so we omit them from our fit.
}
\label{tab:spectra_L2}       
\begin{tabular}{c c c c c c c c c c c c c c c}
\hline\noalign{\smallskip}
 & $n_r$$^{2S+1} L_J$ & $ M_{\rm exp} $ & $ M_{\rm cal} $  &~ $\bar{\vP}^2$ ~& $\sqrt{ \la \vecr^2\ra}$  &~ $f_S$ &~ $\alpha_s$
\\
\hline\noalign{\smallskip}
$\pi_2$ &~$1^1 D_2$        &~ 1.67   ~&~ 1.62 ~&~~0.31~ &~1.25 &~ 0.77  &~ 0.67 \\
           &~$2^1 D_2$          &~ 1.97*   ~&~ 1.98 ~&~~0.42~ &~1.67 &~ ~ &~  \\
           &~$3^1 D_2$          &~ 2.25**   ~&~ 2.28 ~&~~0.54~ &~2.00 &~ ~ &~  \\
\hline\noalign{\smallskip}
$\rho$ &~$1^3 D_1$          &~ 1.57*   ~&~ 1.58 ~&~~0.40~ &~1.13 &~   &~  \\
           &~$2^3 D_1$          &~ 1.91*   ~&~ 1.95 ~&~~0.56~ &~1.53 &~ ~ &~  \\    
           &~$3^3 D_1$          &~ 2.15*   ~&~ 2.21 ~&~~0.71~ &~1.86 &~ ~ &~  \\      
\hline\noalign{\smallskip}
$\rho_2$ &~$1^3 D_2$      &~ ---      ~&~ 1.61 ~&~~0.32~ &~1.22 &~   &~  \\
           &~$2^3 D_2$          &~ 1.94**   ~&~ 1.97 ~&~~0.43~ &~1.65 &~ ~ &~  \\  
           &~$3^3 D_2$          &~ 2.23**   ~&~ 2.27 ~&~~0.55~ &~1.99 &~ ~ &~  \\      
\hline\noalign{\smallskip}
$\rho_3$ &~$1^3 D_3$      &~ 1.69   ~&~ 1.63 ~&~~0.31~ &~1.24 &~   &~  \\
           &~$2^3 D_3$          &~ ---      ~&~ 1.99 ~&~~0.52~ &~1.52 &~ ~ &~  \\  
           &~$3^3 D_3$          &~ 2.30*   ~&~ 2.29 ~&~~0.53~ &~2.01 &~ ~ &~  \\                                 
	\hline\noalign{\smallskip}
	\hline\noalign{\smallskip}
$K_2$ &~$1^1 D_2$        &~ 1.82   ~&~ 1.78 ~&~~0.34~ &~1.18 &~ 0.77  &~ 0.67 \\
           &~$2^1 D_2$         &~ 2.25*   ~&~ 2.13 ~&~~0.53~ &~1.49 &~ ~ &~  \\	
\hline\noalign{\smallskip}
$K^*$ &~$1^3 D_1$          &~ 1.72   ~&~ 1.74 ~&~~0.39~ &~1.12 &~   &~  \\
           &~$2^3 D_1$          &~ ---       ~&~ 2.09 ~&~~0.59~ &~1.44 &~ ~ &~  \\      
\hline\noalign{\smallskip}
$K_2$ &~$1^3 D_2$       &~ 1.77   ~&~ 1.76 ~&~~0.36~ &~1.15 &~   &~  \\
           &~$2^3 D_2$          &~ ---       ~&~ 2.12 ~&~~0.55~ &~1.47 &~ ~ &~  \\    
\hline\noalign{\smallskip}
$K_3^*$ &~$1^3 D_3$      &~ 1.78   ~&~ 1.78 ~&~~0.34~ &~1.19 &~   &~  \\
           &~$2^3 D_3$          &~ ---      ~&~ 2.14 ~&~~0.53~ &~1.50 &~ ~ &~  \\                                          
	\hline\noalign{\smallskip}
	\hline\noalign{\smallskip}
$\eta_{s2}$ &~$1^1 D_2$        &~ 1.84   ~&~ 1.88 ~&~~0.34~ &~1.18 &~ 0.84  &~ 0.67 \\
           &~$2^1 D_2$                 &~ ---      ~&~ 2.18 ~&~~0.53~ &~1.50 &~ ~ &~  \\	       
\hline\noalign{\smallskip}
$\phi$ &~$1^3 D_1$          &~ ---     ~&~ 1.84 ~&~~0.38~ &~1.13 &~   &~  \\
           &~$2^3 D_1$          &~ 2.29  ~&~ 2.14 ~&~~0.57~ &~1.46 &~ ~ &~  \\
\hline\noalign{\smallskip}
$\phi_2 $ &~$1^3 D_2$   &~ ---   ~&~ 1.86 ~&~~0.36~ &~1.16 &~   &~  \\
           &~$2^3 D_2$          &~ ---   ~&~ 2.17 ~&~~0.54~ &~1.49 &~ ~ &~  \\     
\hline\noalign{\smallskip}
$\phi_3 $ &~$1^3 D_3$   &~ 1.85   ~&~ 1.88 ~&~~0.34~ &~1.18 &~   &~  \\
           &~$2^3 D_2$          &~ ---       ~&~ 2.19 ~&~~0.52~ &~1.51 &~ ~ &~  \\    
	\hline\noalign{\smallskip}
	\hline\noalign{\smallskip}                                                      
\end{tabular}
\vspace*{0.1cm}  
\end{table}

The $L \ge 1$ states additional potentials to the $L=0$ potential, (see also Eq.(\ref{eq:av_S_L}))
\beq
\bar{V}_{L=1} 
&=&
\bar{V}_{L=0}  
+ \bar{V}_{ L^2 }  {\bm L}^2 
+ \bar{V}_{ sL } ( {\bm \sigma} \cdot \vL )
\nonumber \\
&& 
+\, \bar{V}_{ (sL)^2 } ( {\bm \sigma}_1 \cdot \vL )( {\bm \sigma}_2 \cdot \vL ) \,.
\label{eq:full_L=1_meson}
\eeq
The $L^2$ and $(sL)^2$ potentials include factors depending on $C_{\calJ_1 \calJ_2}$.
The important new ingredient is the $LS$ potential which depends on the parameter $C_{\calL_i}$. 
The $LS$ forces are as large as the spin-spin forces which, due to the short distance nature, are weakened at $L\ge 1$.
Indeed, most of $1^1P_1$ states are slightly heavier than $1^3P_1$ states at given flavors;
for equal mass flavors, $\la {\bm L} \cdot {\bm S} \ra^{J,J_z}_{L,S} = -S(S+1)$ for $J=L$, see Eq.(\ref{eq:LS}).
To get reasonable $LS$ splittings, we found it necessary to take $C_{\calL_i}$ substantially smaller than $1$.
We take $C_{\calL_i} = 0.5$ in the following. 

In Table.\ref{tab:spectra_L1} we show the spectra for $L=1$ states up to the 1st radial excitations. 
The flavors are $ud, us, ss, uc, sc$, and $cc$, separated by double lines in columns.
The quality of fits are good.
In our fits, the values of $\alpha_s$ are similar to those we chose for the $L=0$ states, while $f_S$ tends to be slightly larger.

Here we mention that the list is not complete.
The scalar mesons such as $a_0$ are omitted from our fit.
For mesons having quarks with unequal masses, the total spin is in general not a good quantum number and hence $S=0$ and $S=1$ states mix for a given total angular momentum $J$.
But we are omitting such mixing terms in the hamiltonian and the results for $S=0$ and $1$ are separately shown.

In Table.\ref{tab:spectra_L2} we show the results of $L=2$ states in the same manner as Table.\ref{tab:spectra_L1}.
The cases for mesons with charm quarks are not displayed, since most of them are not confirmed experimentally.
In our fitting procedures we found that better fits are obtained when we choose the values of $\alpha_s$ considerably smaller than in the $L=0$ and $1$ cases.



\section{Single quark momentum distribution in hadrons}
\label{sec:Q_in}

In this section we use the wavefunctions obtained so far to calculate the single quark momentum distributions in hadrons.

\subsection{Wavefunctions to momentum distributions}
\label{sec:wf_to_Qin}

The single quark distribution in a single hadron with the momentum $\vP_h$ is defined as
($\vP_h$ differs from $\vP$ in the previous sections; the latter has been used for the average quark momentum)
%
\beq
 Q_{\rm in}^{h \vq} (\vp; \vP_h) 
 \equiv 
\la \tilde{\Psi}_{ \vP_h} | n_{ \vq} (\vp) | \tilde{\Psi}_{ \vP_h} \ra
 \,.
 \label{eq:Q_in}
\eeq
Here $\vq = q_s^c$ or $\bar{q}_s^c$ labels a quark (anti-quark) flavor $q (\bar{q})$, spin $s$, and color $c$.
Note that $n_{\vq} $ differs from $q^\dag q \sim a_{\rm p}^\dag a_{\rm p} - a^\dag_{\rm a} a_{\rm a}$ ($a_{\rm p,a}$: annihilation operators for particles and antiparticles)
as we are distinguishing particles and antiparticles; $n_{\vq}$ corresponds to either $a^\dag_{\rm p} a_{\rm p}$ or $a^\dag_{\rm a} a_{\rm a}$.

In this work we neglected the tensor forces and the associated mixing between different $L$. 
We also the mixing between $S=0$ and $S=1$ which occurs when $m_1\neq m_2$.
Within this simplified treatment, our bases for angular momenta can be written as $|J,J_z\ra_{L,S}$.
Hence we write states $h$ specified by $(J,J_z, L,S)$ as\footnote{
Unlike non-relativistic theories, a hamiltonian for relative motions in general is an operator for a given $\vP_h$.
Hence we need to keep the subscript $\vP_h$ for relative wavefunctions.
}
\beq
\hspace{-0.3cm}
 | \tilde{\Psi}_{ \vP_h} \ra 
 = | \vP_h \ra \otimes 
  \sum_F b_h^F | \Psi^F_{ \vP_h} \ra^{ J,J_z}_{L,S} \otimes |F \ra \otimes |C_{\rm s}\ra
\,,
 \eeq
 where $|C_{\rm s}\ra$ is the color singlet state
 \beq
 |C_{\rm s}\ra =  \frac{\, |R \bar{R} \ra + |G \bar{G} \ra + |B \bar{B} \ra \,}{\, \sqrt{\Nc} \,} \,,
 \eeq
 while $F$ is given by combinations of $(u,d,s)$ and $( \bar{u},\bar{d}, \bar{s} )$, e.g., $F= u\bar{u}, d\bar{d}, s\bar{s}, u \bar{d}, ...$, and so on.
 For example, for $| \pi_0\ra$, we have
 \beq
 b_{\pi_0}^{u \bar{u} } = - b_{\pi_0}^{d \bar{d} } = 1/\sqrt{2} \,, 
 ~~~~
 b_{\pi_0}^{s \bar{s} } =  b_{\pi_0}^{u \bar{s} } = \cdots = 0 \,.
 \eeq
 One can readily check
 \beq
 \la F | n_{\vq} (\vp) | F'\ra \propto \delta_{FF'} \,.
 \eeq
 To evaluate $Q_{\rm in}^{h\vq}$, it is convenient to expand the space and spin sectors in the bases $|\vp_1,\vp_2 \ra \otimes |S,S_z\ra$. 
 But before that we first expand in the bases  $|\vecr_1,\vecr_2 \ra $ which we have calculated in the previous sections. 
In coordinate space the wavefunction for a hadron with the momentum $\vP_h$ can be written as 
(in general we compress the vector product as $| a \ra \otimes | b\ra \otimes |c \ra $ into $| a;b;c \ra$ to make the expression compact)
%
\beq
\big[  \tilde{\Psi}^F_{\vP_h } (\vecr_1,\vecr_2) \big]^{ J,J_z}_{L,S,S_z} 
  &\equiv &
  \la \vecr_1, \vecr_2; S,S_z| \vP_h;  \tilde{\Psi}^F_{ \vP_h} \ra^{ J,J_z}_{L,S} 
  \nonumber \\
  &= &
  \frac{\, \rme^{\rmi \vP_h \cdot \vR_h } \,}{\, \sqrt{V} \,} \big[ \Psi^F_{\vP_h} (\vecr) \big]^{ J,J_z}_{L,S,S_z}  \,,
\eeq
where ($c_1+c_2=1$)
\beq
\vR_h = c_1 \vecr_1 + c_2 \vecr_2 \,,~~~~~ \vecr = \vecr_1 -  \vecr_2 \,,
\eeq
or $\vecr_1 = \vR_h + c_2 \vecr$ and $\vecr_2 = \vR_h - c_1 \vecr$.
For $\vP_h\neq 0$, the choice of $(c_1,c_2)$ depends on hadrons as we will discuss shortly in Sec.\ref{sec:approx_moving_hadrons}.
The Fourier transform leads to (we suppress the indices for the moment)
\beq
\tilde{\Psi}_{\vP_h} (\vp_1,\vp_2) 
&=& \int_{\vecr_1, \vecr_2} \rme^{ - \rmi (\vp_1 \vecr_1 + \vp_2 \vecr_2) } \tilde{\Psi}_{\vP_h} (\vecr_1,\vecr_2) \nonumber \\ 
&=& \frac{\, (2\pi)^3 \,}{\, \sqrt{V} \,} \delta(\vp_1 +\vp_2-\vP_h) \Psi_{\vP_h} (\vp_r) \,,
\eeq
where
\beq
\vP_h = \vp_1 + \vp_2 \,,~~~~~ \vp_r = c_2 \vp_1 - c_1 \vp_2 \,.
\eeq
We note that the $\vP_h=0$ case in the previous sections corresponds to $\vp_1=-\vp_2 = \vp_r$.
The normalization is $\int_{\vp_r} | \Psi_{\vP_h} (\vp_r)|^2=1$.

Now one can write
\beq
 | \tilde{\Psi}_{ \vP_h} \ra 
 &=& 
   \sum_{F,S_z} b_h^F 
   \int_{\vp_1,\vp_2}
  \big[   \tilde{\Psi}^F_{\vP_h } (\vp_1,\vp_2) \big]^{ J,J_z}_{L,S,S_z} 
  \nonumber \\
&& ~~ \times    | \vp_1,\vp_2 \ra \otimes |S,S_z\ra \otimes |F \ra \otimes |C_{\rm s}\ra
\,,
 \eeq
 In this base the occupation number operator can be evaluated readily.
 Assuming that we pick up the particle 1, we obtain
\beq
&&
 \la \vp'_1,\vp'_2; S_z' ;F';C_{\rm s} | n_{\vq} (\vp_1) |   \vp''_1,\vp''_2; S''_z ;F'';C_{\rm s} \ra_S
\nonumber \\
&& = (2\pi)^9 \delta(\vp_1-\vp'_1) \delta(\vp'_1-\vp''_1) \delta(\vp'_2-\vp''_2) \delta_{F'F''} \frac{\, N_{\vq SF'}^{S'_z S''_z} \,}{\, \Nc \,} \,.
\nonumber \\
\eeq
The factor $1/\Nc$ reflects that a single color in a meson can be found with the probability $1/\Nc$.
Now one can write
\beq
 && 
 \hspace{-0.2cm}
 Q_{\rm in}^{h \vq} (\vp_1; \vP_h) 
 \nonumber \\
&&  \hspace{-0.2cm}
= 
 \sum_{F,S'_z,S''_z} \frac{\, |b_h^F |^2 \,}{\, \Nc \,} \, N_{\vq SF}^{S'_z S''_z}
\nonumber \\
&&
~ \times   \int_{\vp'_2}
  \tilde{\Psi}^F_{\vP_h} (\vp_1,\vp'_2)^{ J,J_z}_{L,S,S_z'}   \tilde{\Psi}^{*F}_{\vP_h} (\vp_1,\vp'_2)^{ J,J_z}_{L,S,S''_z}
  \nonumber \\
&&  \hspace{-0.2cm}
=   \sum_{F,S'_z,S''_z} \frac{\, |b_h^F |^2 \,}{\, \Nc \,} \, N_{\vq SF}^{S'_z S''_z}
\nonumber \\
&&
~ \times  
 \Psi^F_{\vP_h} (\vp_1 -c_1 \vP_h)^{ J,J_z}_{L,S,S_z'}  \Psi^{*F}_{\vP_h} (\vp_1-c_1 \vP_h)^{ J,J_z}_{L,S,S''_z} \,.
\eeq
For a hadron $h = h^{J J_z}_{LS}$ with the angular momentum $(J,J_z,L,S)$, the momentum distribution is generally anisotropic in momentum space.
As we see later, we are interested in the distribution after summing over the quantum number $J_z$.
Then the expression can be further simplified.
Writing $\vp_r = \vp_1 - c_1 \vP_h$, $p_r = |\vp_r|$, and $\hat{\vp}_r = \vp_r/p_r$,
%
\beq
\hspace{-0.cm}
&&\la \vp_r ;S, S_z | \Psi^F_{ \vP_h} \ra^{ J,J_z}_{L,S} 
\nonumber \\
&&= \phi^{F,L}_{\vP_h} (p_r) 
	 \sum_{L_z } C^{J,J_z}_{L,L_z; S,S_z } Y_L^{L_z} (\hat{\vp}_r) \,,
\eeq
with $Y_L^{L_z}$ being the spherical harmonic function,
and
using the closure relation for the Clebsch-Gordan coefficients
\beq
\sum_{J_z} C^{J,J_z}_{L,L_z; S,S_z } \big( C^{J,J_z}_{L,L'_z; S,S'_z } \big)^*
 = \delta_{S_z,S'_z} \delta_{L_z,L_z'} \,,
\eeq
and the closure relation 
\beq
\sum_{L_z} Y_L^{L_z} (\hat{\vp}_r) Y_L^{L_z} (\hat{\vp}_r)^*=(2L+1)/4\pi\,,
\eeq
we obtain the formula for $Q_{\rm in}$ summed over $J_z$.
Summing hadrons with $h = h^{J}_{LS}$,
\beq
 \sum_{J_z} 
 Q_{\rm in}^{h \vq} (\vp_1; \vP_h) 
= 
\frac{\, (2L+1) \,}{\, \Nc \,} 
  \sum_{F,S_z}  |b_h^F |^2  \, N_{\vq SF}^{S_z }\,
 \frac{\, | \phi^{F,L}_{\vP_h} (p_r) |^2 \,}{\, 4\pi \,} \,.
 \nonumber \\
 \label{eq:Q_in_Jz_sum}
\eeq
The function $\phi_{\vP_h}^{F,L} (p_r)$ is isotropic in momentum space, and
is normalized as $\int_0^\infty \rmd p_r \, p_r^2 | \phi_{\vP_h}^{F,L} (p) |^2 =(2\pi)^3$.

Finally we look at some examples of $N_{\vq SF}^{S_z}$.
For instance, $\rho_0^{S_z=1}$ states yield $b_{\rho_0}^{u \bar{u} } = - b_{\rho_0}^{d \bar{d} } = 1/\sqrt{2} $, and the spin state $|S=1,S_z=1\ra = |\up \up\ra$,
resulting in
\beq
\big( N_{u^R_\up} \big)_{ S=1, F = u\bar{u}  } ^{S_z = 1 }
=  \la u^R_\up \bar{u}^R_\up | N_{u_\up^R} | u^R_\up \bar{u}^R_\up \ra = 1 \,. 
\eeq
Another example is $\rho_0^{S_z=0}$, for which 
\beq
\big( N_{u^R_\up} \big)_{ S=1, F = u\bar{u}  } ^{S_z = 0 }
=  \la u^R_\up \bar{u}^R_\down | N_{u_\up^R} | u^R_\up \bar{u}^R_\down \ra/ (\sqrt{2})^2 = 1/2 \,. 
\nonumber \\
\eeq
In both cases $F=d\bar{d}$ component is zero.
Some lists for the color-flavor-spin factors are given in Appendix.\ref{sec:color_flavor_spin_indices}.

\subsection{Distributions in hadrons at rest }
\label{sec:distributions_at_Ph0}

\begin{figure}[bt]
\vspace{-0.cm}
\begin{center}	
	\includegraphics[width=8.5cm]{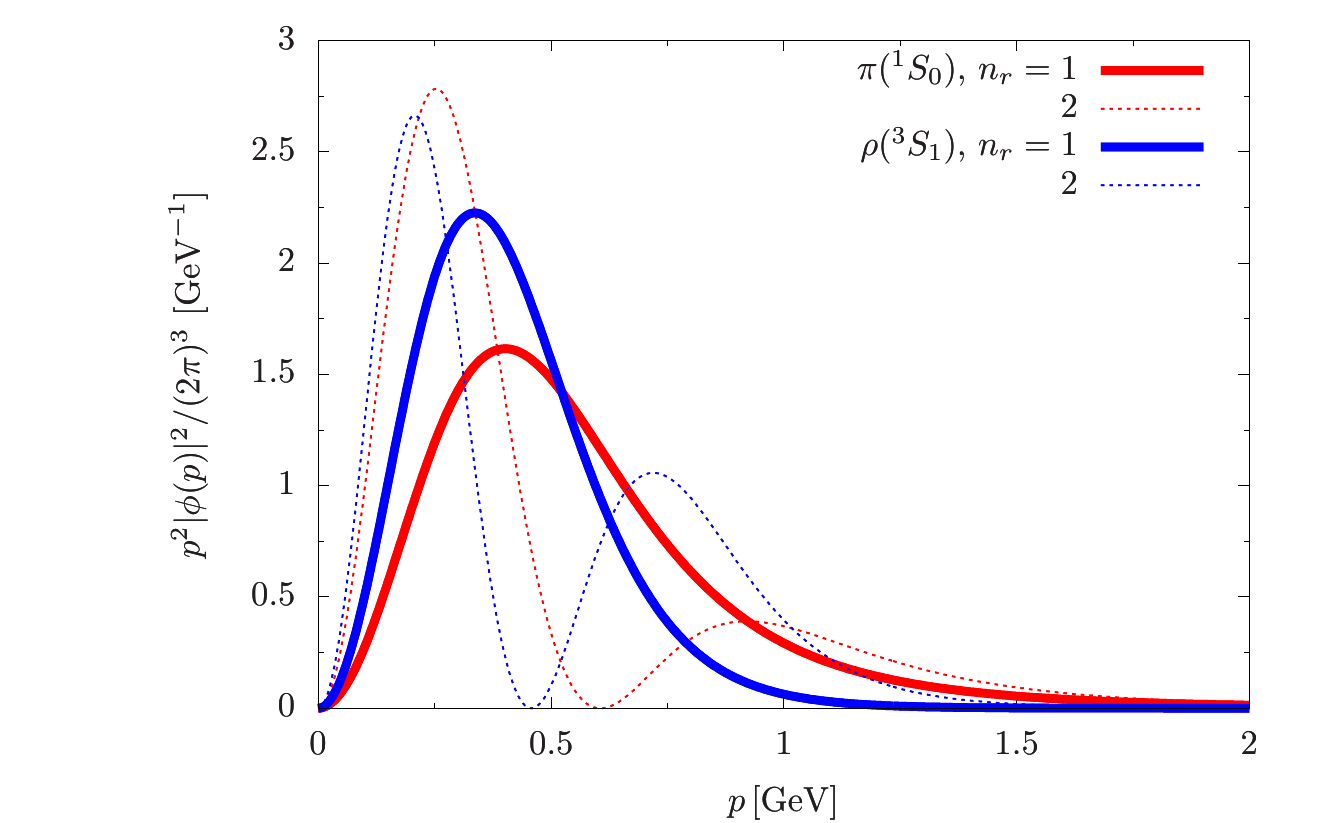}
	\includegraphics[width=8.5cm]{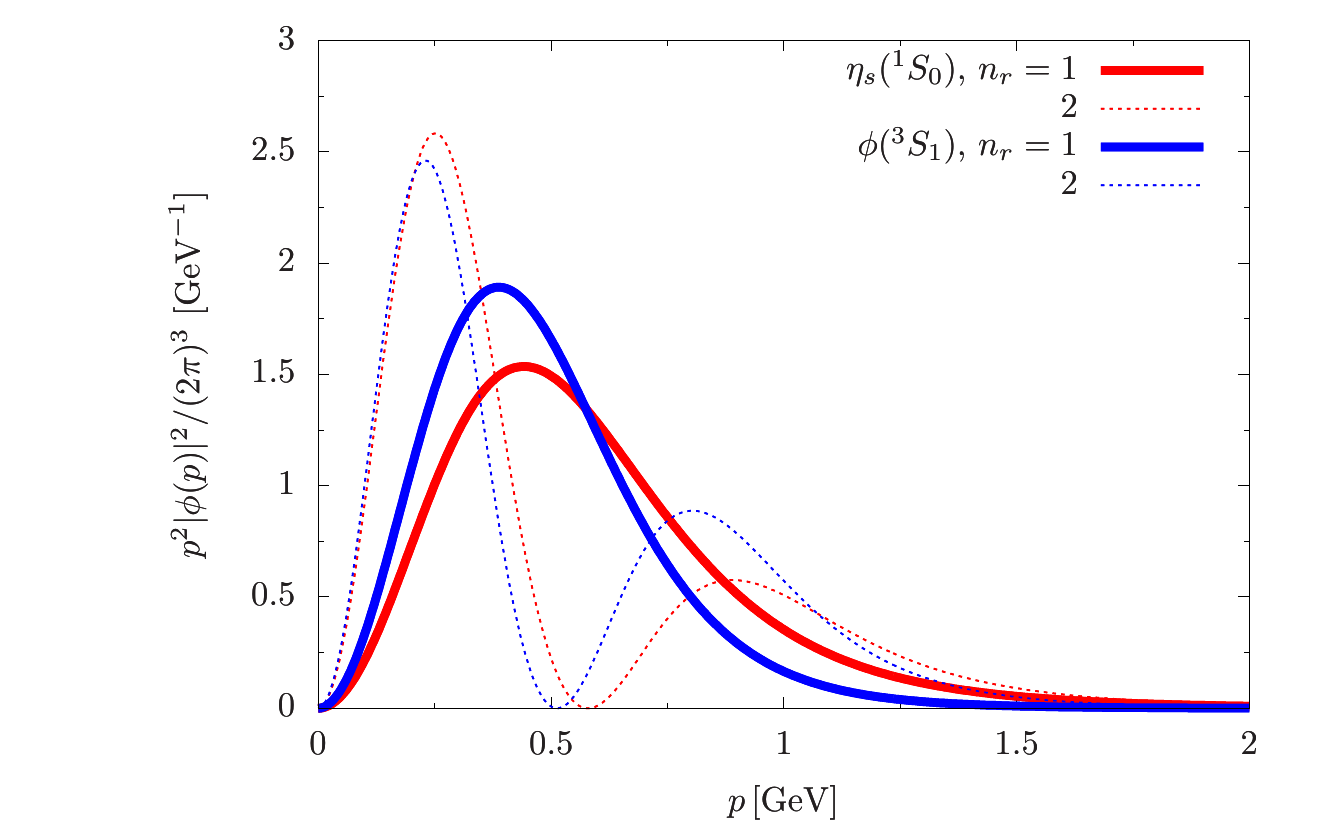}

		\vspace{-0.2cm}
\caption{  The single quark momentum distribution $p^2 |\phi^{F,L} (p)|^2/(2\pi)^3\, [{\rm GeV}^{-1}]$ for $\pi\, (^1S_0)$, $\rho\, (^3S_1)$ states (upper panel) and $\eta_s\, (^1S_0)$, $\phi\, (^3S_1)$ states (lower panel). The radial excitations with $n_r=2$ are also shown in thin lines. The distributions for $s$-quarks are shifted slightly to higher momenta than for $ud$-quarks.
		 }
		 		\vspace{-0.4cm}
		 \label{fig:p2_phi^2_ud_ss}
\end{center}		
		
\end{figure}

\begin{figure}[bt]
\vspace{-0.cm}
\begin{center}	
	\includegraphics[width=8.0cm]{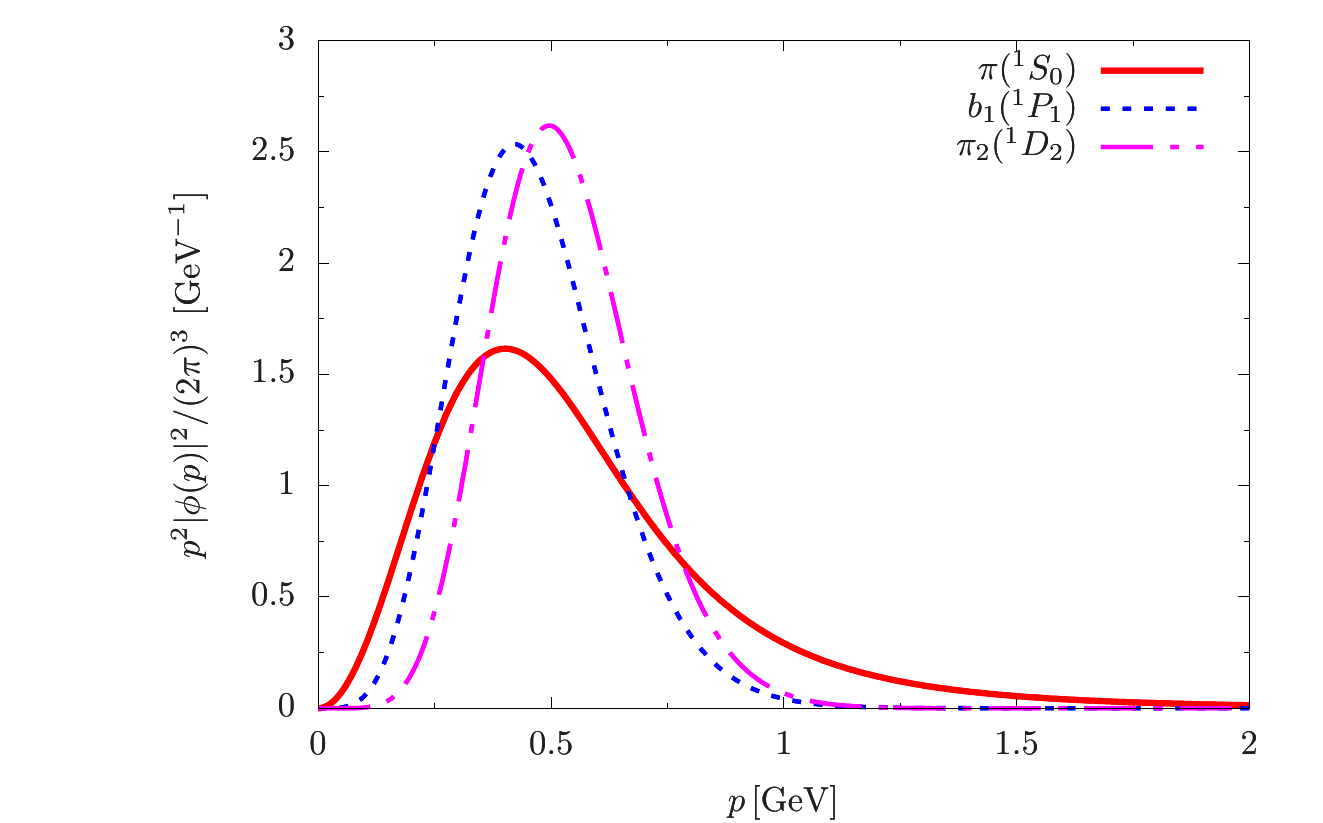}
	
		\vspace{-0.2cm}
\caption{  The momentum distribution $p^2 |\phi(p)|^2\, [{\rm GeV}^{-1}]$ for the $^{2s+1}L_j = \,^1L_L$ states, $\pi\, (^1S_0)$, $b_1\, (^1P_1)$, and $\pi_2\, (^1D_2)$ states.
		 }
		 		\vspace{-0.4cm}
		 \label{fig:p2_phi_2_ud_higher_L}
\end{center}		
		
\end{figure}

Now we examine the quark momentum distributions in a hadron with $h = h^{J J_z}_{LS}$  at rest ($\vP_h=0$).
Shown in Fig.\ref{fig:p2_phi^2_ud_ss} are the momentum distributions $p^2 |\phi(p)|^2\, [{\rm GeV}^{-1}]$ for $\pi\, (^1S_0)$, $\rho\, (^3S_1)$ states (upper panel) 
and $\eta_s\, (^1S_0)$, $\phi\, (^3S_1)$ states (lower panel). The radial excitations with $n_r=2$ are also shown in thin lines. 
Due to the spin-spin attractions, $\pi$ is more compact than $\rho$, and hence tend to have higher momenta.
In general, the $n_r$-th radial excitation has the distributions with $n_r$ peaks distributed from low to high momenta.
We note that higher radial excitations contain more low momentum components than lower radial excitations.
This is likely due to the broader spatial structure in higher radial excitations.
The distributions for $s$-quarks are shifted slightly to higher momenta than for $ud$-quarks.

Shown in Fig.\ref{fig:p2_phi_2_ud_higher_L} is the same as Fig.\ref{fig:p2_phi^2_ud_ss} but for different states with $^{2s+1}L_j = \,^1L_L$ states, 
$\pi\, (^1S_0)$, $b_1\, (^1P_1)$, and $\pi_2\, (^1D_2)$ states.
Unlike the radial excitations, with a greater $L$ the distributions simply get shifted to higher momenta,
and soft momentum components become smaller.
The standard centrifugal barrier effects disfavor soft momentum components to develop.

\subsection{Approximations for moving hadrons}
\label{sec:approx_moving_hadrons}

To examine the properties of matter in hot and dense media, we need hadrons at finite momenta.
At finite $\vP_h$, several new elements enter:
(i) in the relativistic treatments, the relative wavefunctions as functions of $\vp_r$ generally also depend on $\vP_h$ (the center of mass motion does not fully decouple);
and (ii) we have to make an appropriate choice for $(c_1,c_2)$.
The question is how the presence of a finite $\vP_h$ affects our eigenvalue problem.
Since our eigenvalue problems were not manifestly covariant,
we compromise with some approximate treatments. 
For spectra, we simply assume $E(\vP_h=0) = m_h \rightarrow E(\vP_h) = \sqrt{m_h^2 + \vP_h^2}$,
without explicitly demonstrating it in our model.
Meanwhile some extra discussions are needed for the wavefunctions.
At this point we choose a specific combination of $(c_1,c_2)$ affect 
our eigenvalue problems including terms like $E_{1,2}= \sqrt{m_{1,2}^2 + \vp_{1,2}^2 \,}$.
One possible choice is to take 
\beq
c_{ih} = \frac{\, \bar{E}_i \,}{\, \bar{E}_1 + \bar{E}_2 \,} = \frac{\, \bar{E}_i \,}{\, \bar{E}_{1+2} \,} 
 \,,~~~~ \bar{E}_i = \sqrt{ m_i^2 + \bar{\vP}^2 } \,,
\label{eq:c_i}
\eeq
where the subscript $h$ is attached to emphasize that $\bar{\vP}^2$ depend on a meson being considered.
The $c_{ih}$ gives $1/2$ in the equal mass case $(m_1=m_2)$ or in the relativistic limit ($\bar{\vP}^2/m_i^2 \rightarrow \infty$),
while leads to $m_i/(m_1+m_2)$  in the non-relativistic limit ($\bar{\vP}^2/m_i^2 \rightarrow 0$).
Substituting $\vp_1 = c_{1h} \vP_h + \vp$ and $\vp_2 = c_{2h} \vP_h - \vp$, and expanding in powers of $\vP_h$ around $\vP_h=0$,
the $E_{1+2} $ term commonly appearing in our eigenvalue problem looks
\beq
\hspace{-0.5cm}
E_{1+2} (\vP_h)
= E_{1+2}^{\vP_h=0}
&+& \bigg( \frac{c_{1h} }{\, E_1 \,} - \frac{c_{2h} }{\, E_2 \,} \bigg) \bigg|_{\vP_h=0} \vp \cdot \vP_h 
\nonumber \\
&+& 
\bigg( \frac{c_{1h}^2 }{\, E_1 \,} + \frac{c_{2h}^2 }{\, E_2 \,} \bigg) \bigg|_{\vP_h=0} \frac{\,  \vP_h^2 \,}{\, 2 \,} 
+ \cdots
\eeq
Expanding $E_i$ in the denominator around $ \bar{E}_i$, the cross term $\vp \cdot \vP_h$ vanishes, and
\beq
E_{1+2} (\vP_h)
\simeq 
E_{1+2}^{\vP_h=0}
+
 \frac{\,  \vP_h^2 \,}{\, 2 \big( \bar{E}_1 +  \bar{E}_2 \big) \,} 
 + \cdots \,.
\eeq
%
The way  $\vP_h^2$ appears is similar to what we would have for non-relativistic approximation for mesons, $E(\vP_h) \simeq m_h + \vP_h^2/ 2m_h + \cdots$.
This motivates us to use $c_i$'s  in Eq.(\ref{eq:c_i}), and the choice leads to the correct limiting behaviors in relativistic and non-relativistic limits.
The quality of our approximation may be tested by comparing the above corrections in spectra with what we expect from the Poincare invariance, $\vP_h^2/ 2m_h $.
For $\vP_h \ll \bar{E}_{1,2}$, we expect that it is a reasonable approximation to expand
 \beq
 \Psi_{\vP_h} (\vp) \simeq  \Psi_{\vP_h=0} (\vp) + \vP_h \cdot \partial_{\vP_h} \Psi (\vp) + \cdots  \,,
 \label{eq:approx_Ph}
\eeq
as a weak external momentum does not affect internal wavefunctions characterized by hard momenta.

\section{Meson resonance gas}
\label{sec:HRG}

Using the wavefunctions obtained so far, we compute the single quark momentum distributions $f^T_{\vq} (\vp)$ 
at a given temperature $T$ in a meson resonance gas (MRG).
The distribution is compared to what we would obtain from percolated quarks extended over space.

\subsection{Quark occupation probability: formula}
\label{sec:quark_occ_formula}

The $f_\vq^T$ is computed by simply summing up the quark momentum distribution from each meson $h$ \cite{Kojo:2021ugu,Kojo:2021hqh},
\beq
f_\vq^T (\vp) = \sum_h \int_{\vP_h} n_h^T (\vP_h) Q_{\rm in}^{h \vq} (\vp;\vP_h)  \,,
\label{eq:fq_nh_Qin}
\eeq
where we assume the hadron $h$ with the energy $E_h (\vP_h) = \sqrt{ m_h^2 + \vP_h^2}$ obeys the Bose-Einstein distribution, $n_h^T (\vP_h) = [\, \rme^{E_h(\vP_h)/T}-1 \,]^{-1} $.
The function $Q_{\rm in}^{h \vq} (\vp;\vP_h)$ specifies the distribution of a quark with the quantum number $\vq$ and momentum $\vp$,
in a hadron $h$ with the momentum $\vP_h$.
Using the formula Eq.(\ref{eq:Q_in}) for a single quark momentum distribution, we obtain
\beq
\hspace{-0.4cm}
f_\vq^T (\vp) 
&=& \sum_{h=h^J_{LS} }  \int_{\vP_h} n_h^T (\vP_h ) 
\nonumber \\
&&  \times
 \sum_{F,S_z} \frac{\, |b_h^F |^2 \,}{\, \Nc \,} \, N_{\vq SF}^{S_z } (2L+1)\,
  \frac{\, | \phi^{F,L}_{\vP_h} (p_r) |^2 \,}{\, 4\pi \,} \,,
 \label{eq:fq^T}
\eeq
where $p_r = |\vp_r|$ with $\vp_r = \vp - c_{1h} \vP_h$, and
the sum over $h=h^J_{LS}$ excludes the summation over $J_z$ which has been taken into account in the last factors of Eq.(\ref{eq:fq^T}).

To treat this expression numerically, it is convenient to use $\vk = \vp - c_{1h_F} \vP_h$ as an integration variable
where we update $h \rightarrow h_F$ to emphasize that $c_{1h}$ depends on the flavors $F$ in a hadron $h$; for instance $u$- and $s$-quarks in $\eta$ or $K$ have different $c_{1h}$.
Using the approximation in Eq.(\ref{eq:approx_Ph}), 
\beq
\hspace{-0.4cm}
f_\vq^T (\vp) 
&\simeq & \sum_{h^J_{LS}, F} 
	\frac{1}{\, c_{1h_F}^3 \,}\int_{\vk} n_h^T \bigg( \frac{\, \vp-\vk \,}{c_{1h_F} }  \bigg) 
\nonumber \\
&&  \times
	 \sum_{S_z} \frac{\, |b_h^F |^2 \,}{\, \Nc \,} \, N_{\vq SF}^{S_z } (2L+1)\,
	 \frac{\, | \phi^{F,L}_{\vP_h=0} (k) |^2 \,}{\, 4\pi \,} \,.
\label{eq:appro_fq}
\eeq
The results presented below are based on this formula.
\subsection{Numerical results }
\label{sec:numerical}

\begin{figure}[bt]
\vspace{-0.cm} 
\begin{center}	
	\includegraphics[width=8.5cm]{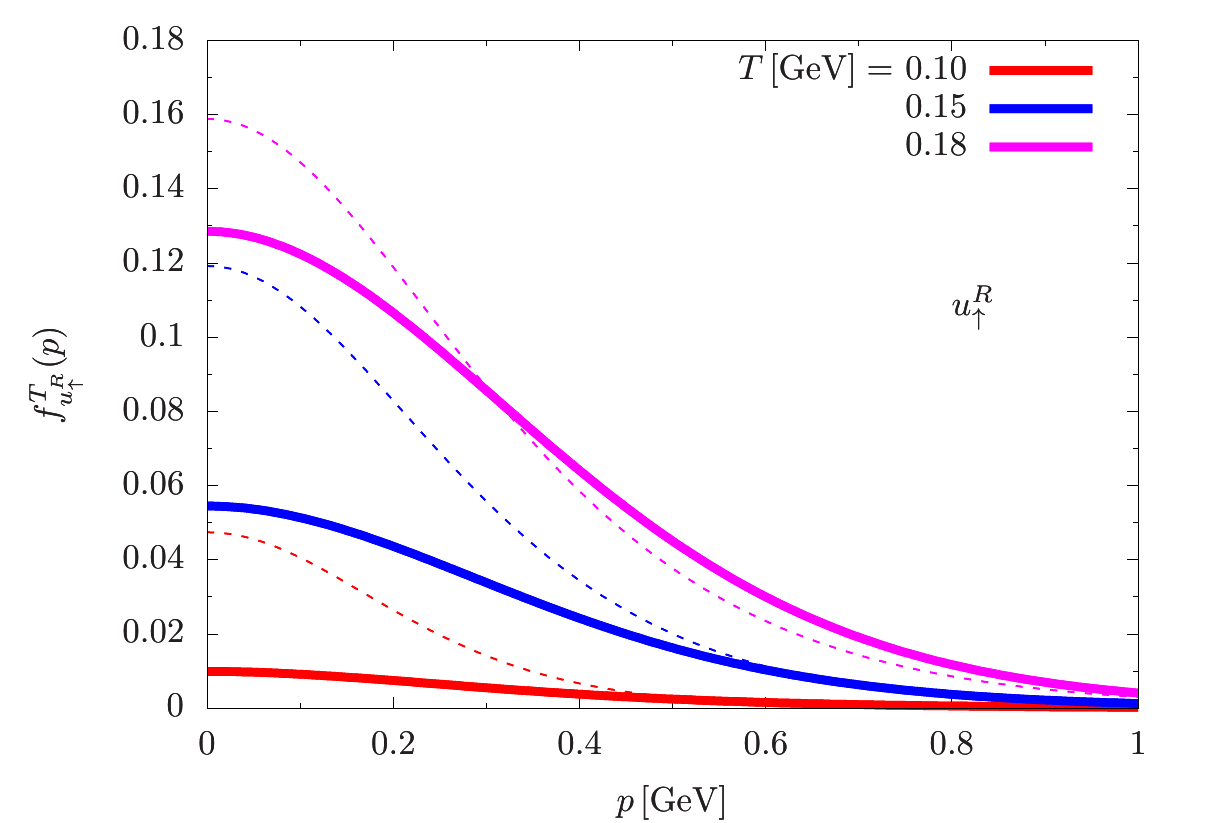}
	\includegraphics[width=8.5cm]{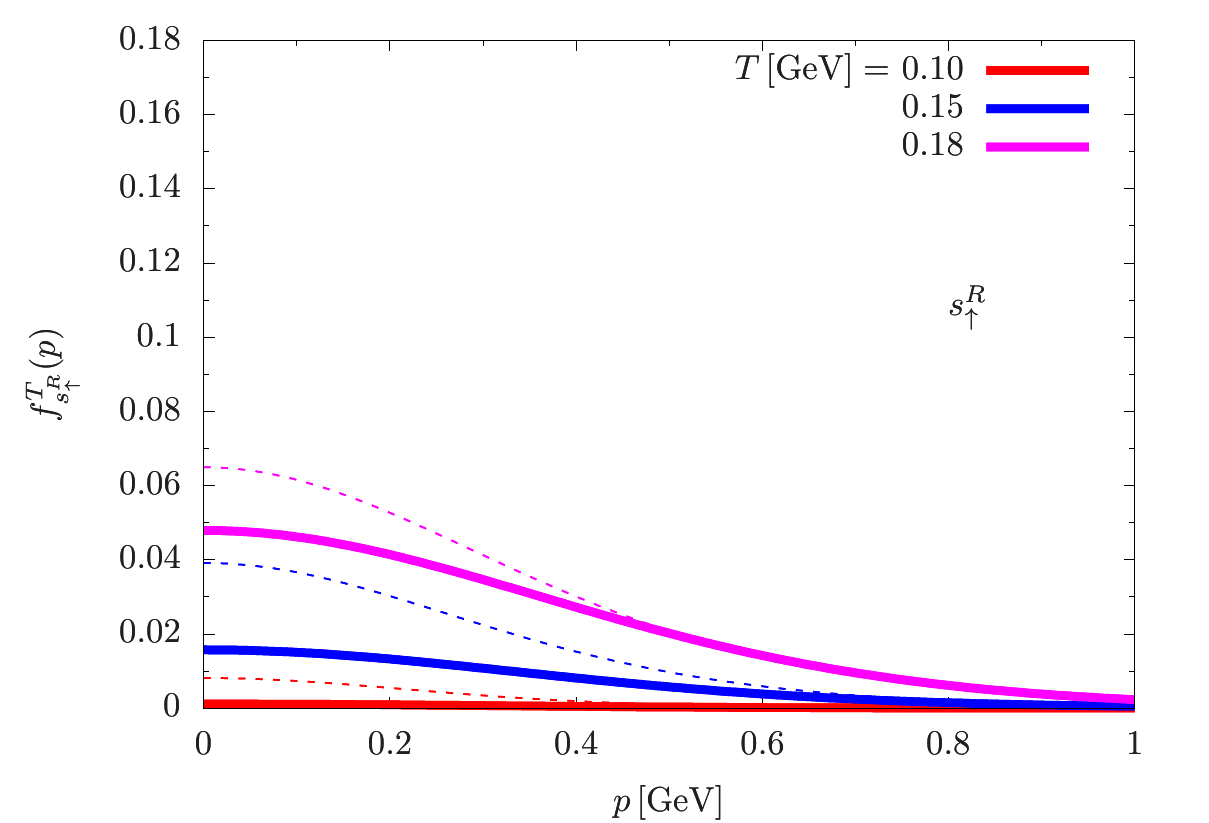}

		\vspace{-0.2cm}
\caption{  The occupation probabilities of single quark states, $f^T_{\vq} (p)$, in our meson resonance gas model (bold lines) and in an ideal constituent quark gas (thin lines). 
The temperatures are $T=0.10, 0.15$, and $0.18$ GeV. Results for the $u^R_\up$ (upper panel) and $s^R_\up$ (lower panel) states are shown. All the mesons up to $L=2$ and $n_r=3$ are included.
		 }
		 		\vspace{-0.4cm}
		 \label{fig:fq_ud_ss}
\end{center}		
		
\end{figure}

\begin{figure}[bt]
\vspace{-0.cm} 
\begin{center}	
	\includegraphics[width=8.5cm]{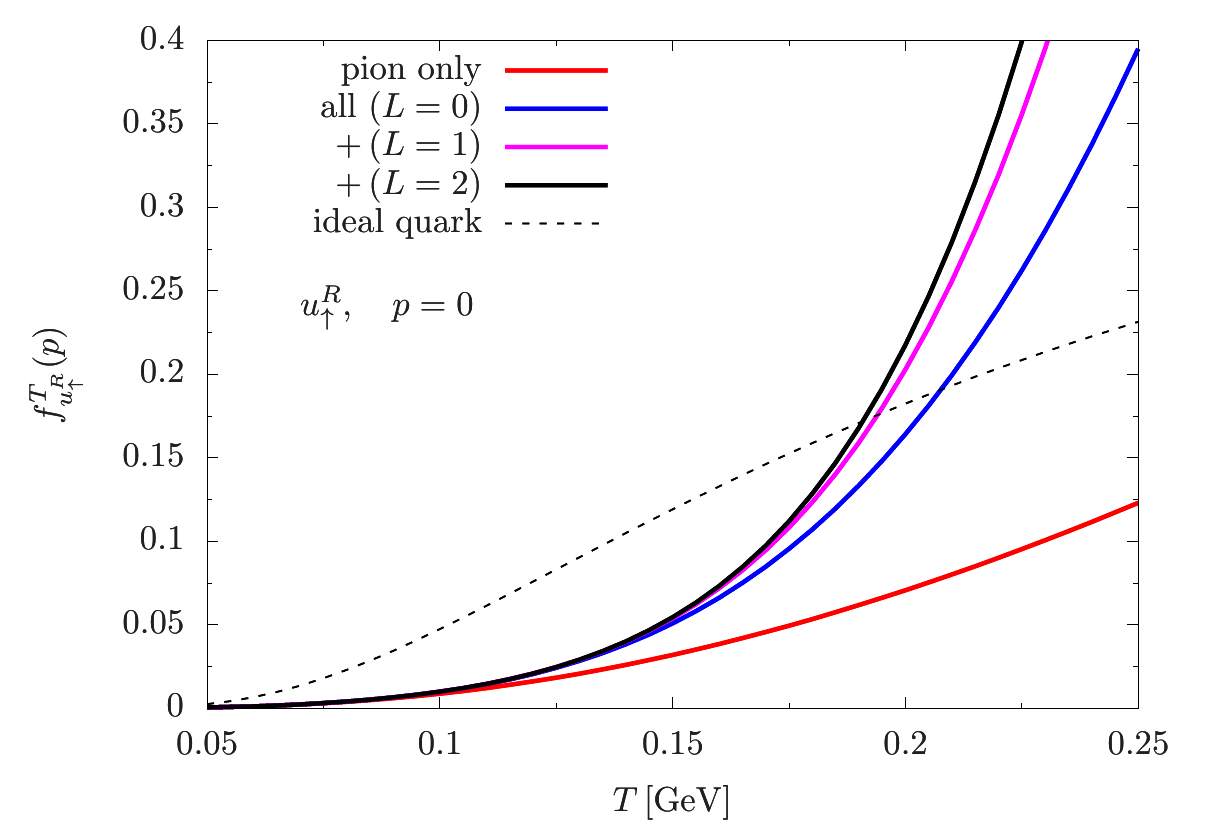}
	\includegraphics[width=8.5cm]{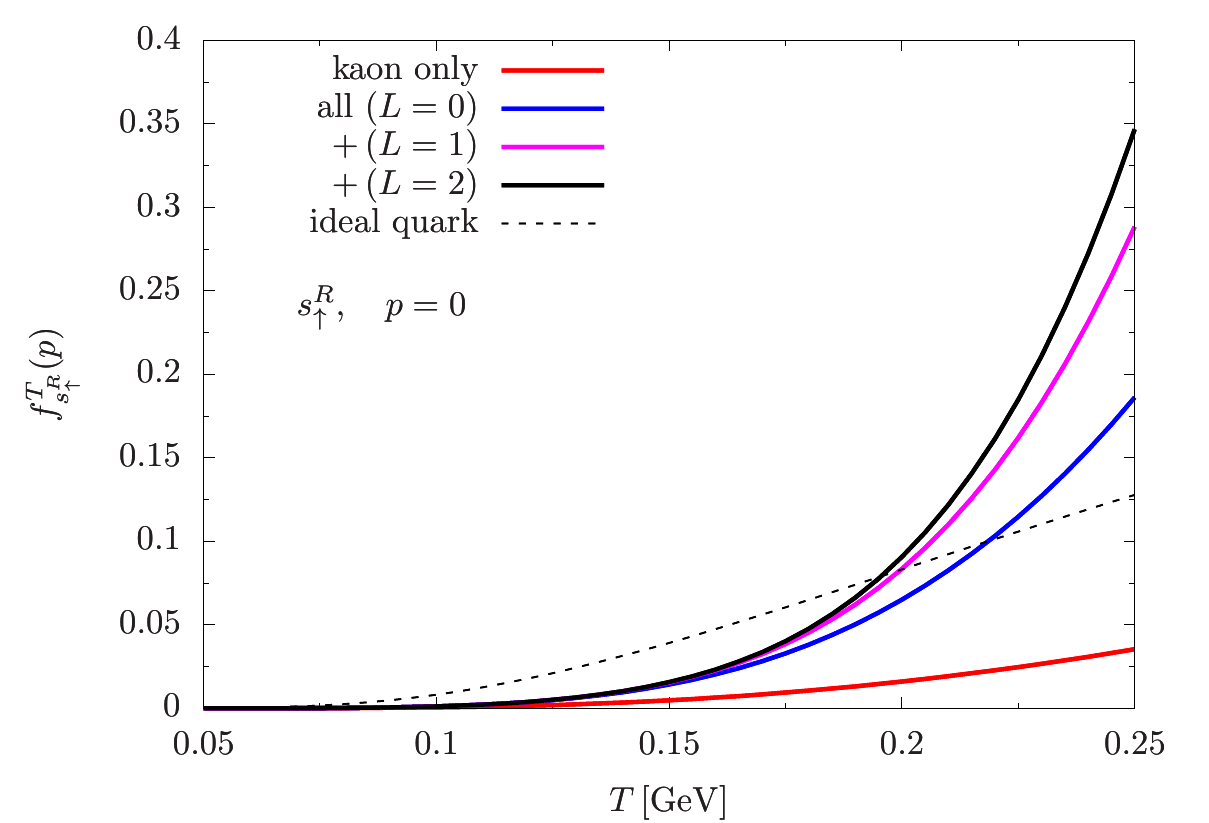}

		\vspace{-0.2cm}
\caption{  
The occupation probabilities of $u^R_\up$ (upper panel) and $s^R_\up$ (lower panel) at $p=0$ as functions of temperature.
The dashed lines are used for $f_{\vq}^{\rm id}$.
Solid curves are used for $f_{u^R_\up}^T$. 
In the upper (lower) panel, we sequentially add pions (kaons), radial excitations up to $n_r=3$ for the $L=0, 1$, and 2. 
The $ud$, $us$, $ss$ type mesons are included.
		 }
		 		\vspace{-0.4cm}
		 \label{fig:fq_ud_ss_T_dep}
\end{center}		
		
\end{figure}

Now we numerically evaluate the integral in Eq.(\ref{eq:appro_fq}).
We compare $f_{\vq}^T$ in our MRG with the distribution for an ideal quark gas, 
$f_\vq^{\rm id} = [\rme^{E_\vq (\vp)/T}+1]^{-1}$, with the constituent quark masses same as used in our MRG, $M_u=0.30$ GeV and $M_s= 0.48$ GeV.
We regard the occupation probability of such a quark gas as that in a QGP, in the sense that quark states can be extended over space.

First we recall that the beginning of the crossover is $T\sim 150$ MeV in QCD where the HRG descriptions begin to break down.
Another point to be kept in mind is that our model does not include baryons.
(Meanwhile, the present results may be used to get some insights for two-color QCD where diquark baryons degenerate with mesons.)
Keeping these insufficiencies in mind, we examine some trends in the following.

Shown in Fig.\ref{fig:fq_ud_ss} are the occupation probabilities of single quark states, 
$f^T_{\vq} (p)$, in our MRG model (bold lines) 
and, $f_\vq^{\rm id}$, in a QGP (thin lines). 
The temperatures are $T=0.10, 0.15$, and $0.18$ GeV. 
Results for the $u^R_\up$ (upper panel) and $s^R_\up$ (lower panel) states are shown. All the mesons up to $L=2$ and $n_r=3$ are included.

At low temperatures, the $f_\vq^T$ is lower than $f_\vq^{\rm id}$ at low momenta.
In the latter, the confining effects are absent so that quark wavefunctions can be widely extended, occupying low momentum states.
As temperature increases, the number of hadrons increases drastically and quarks inside of such hadrons radically occupy low momentum states.
The occupation probability at low momenta eventually exceeds $f_\vq^{\rm id}$.
Such temperature dependent evolution can be seen in Fig.\ref{fig:fq_ud_ss_T_dep}, 
where hadronic contributions are added sequentially to $f_\vq^T(p)$ at $p=0$.
In the upper (lower) panel for $f_{u^R_\up}$ ($f_{s^R_\up}$), only pions (kaons) with $n_r=1$ are included for the lowest curve;
the next solid curve includes all $L=0$ hadronic states up to $n_r =3$, where $ud$, $us$, $ss$ type mesons are taken into account.
We continue such adding to $L=2$.
Around $T\sim 0.15$ GeV, the contributions from the radial excitations are substantial, as they include considerable $p=0$ components (Fig.\ref{fig:p2_phi^2_ud_ss}).
The excess $f_{\vq}^T$ over $f_\vq^{\rm id}$ happens around $T\sim 0.2$ GeV.
We emphasize that our MRG underestimates the true occupation probability which should be enhanced by baryons.
If we include baryons, the $f_{\vq}^T$ should exceed $f_{\vq}^{\rm id}$ at lower temperature, perhaps somewhere around $0.15$-$0.18$ GeV.

\begin{figure}[bt]
\vspace{-0.cm} 
\begin{center}	
	\includegraphics[width=8.5cm]{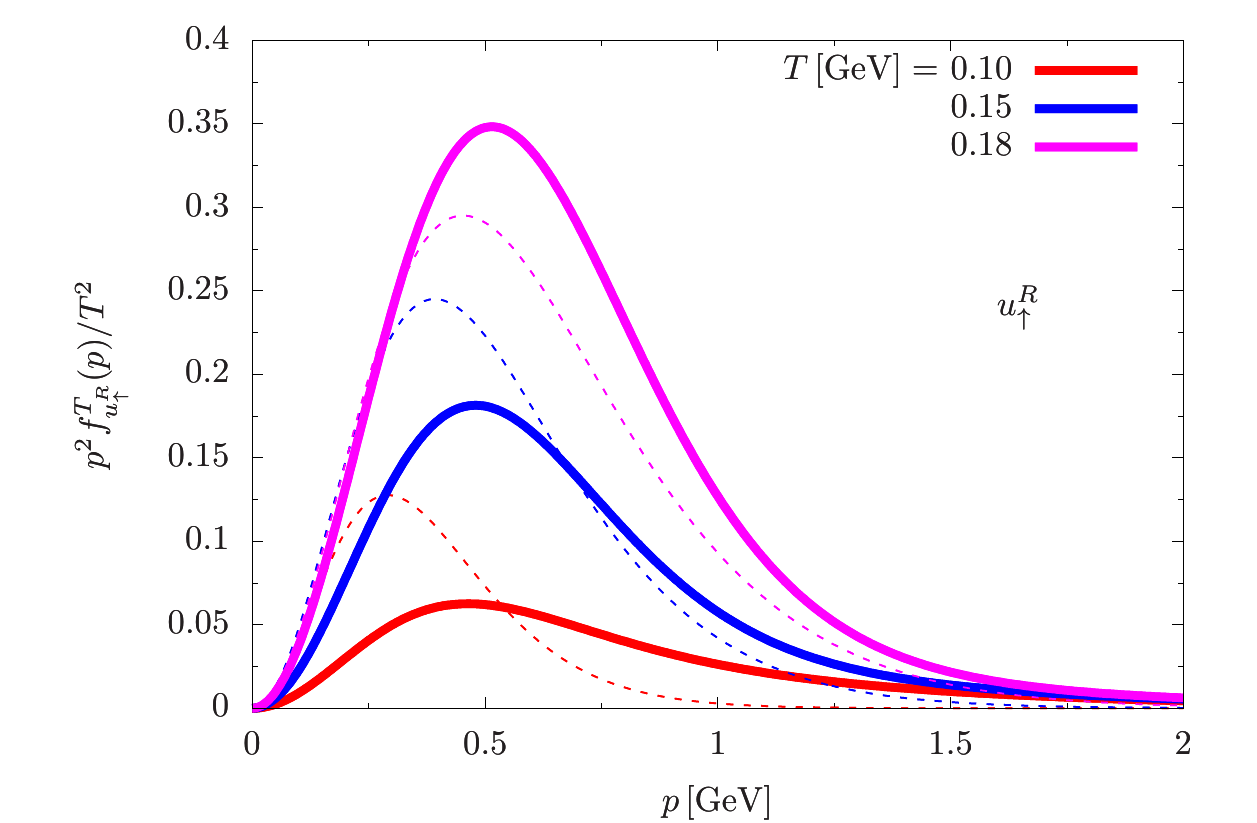}
	\includegraphics[width=8.5cm]{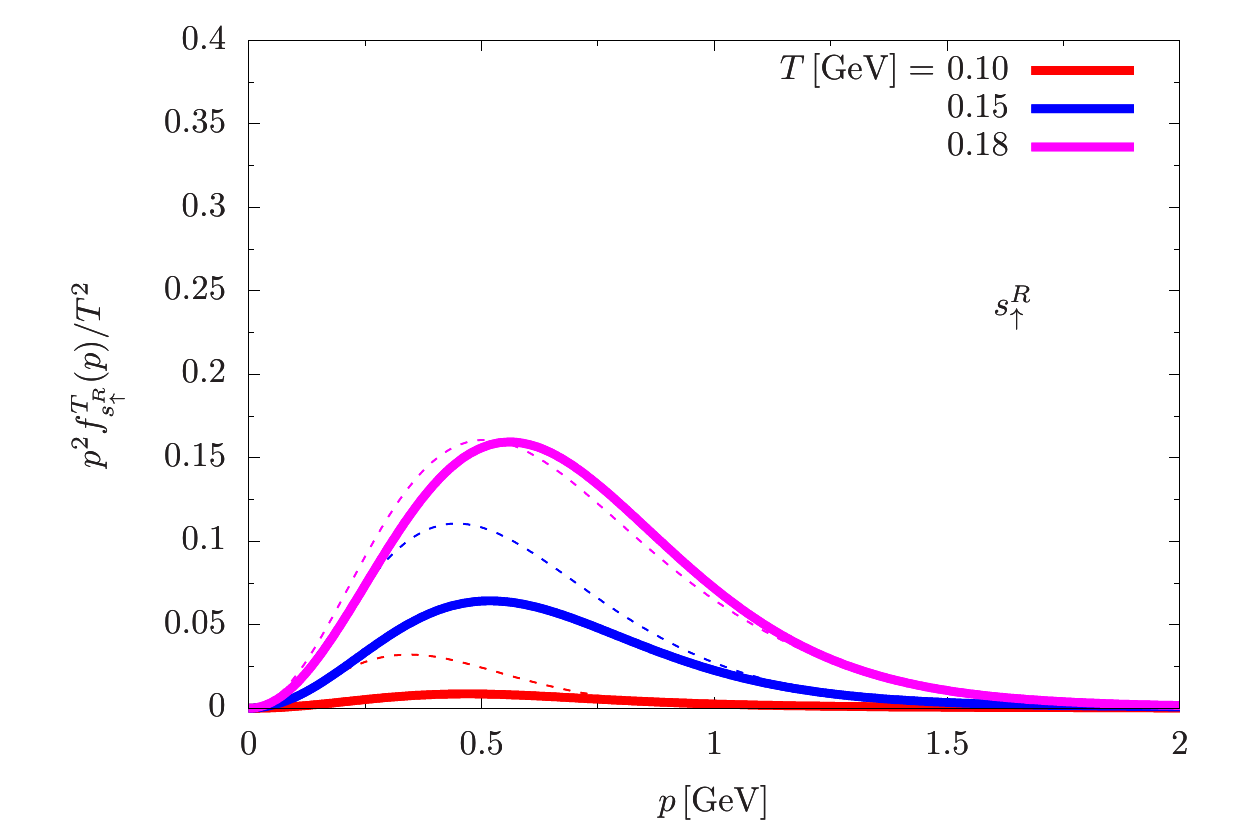}

		\vspace{-0.2cm}
\caption{  The same as Fig.\ref{fig:fq_ud_ss} but the factor $p^2/T^2$ is multiplied.
		 }
		 		\vspace{-0.4cm}
		 \label{fig:fq_ud_ss_p2}
\end{center}		
		
\end{figure}

To examine which momentum states are typical, it is useful to multiply the phase space factor $4\pi p^2$.
Shown in Fig.\ref{fig:fq_ud_ss_p2} are the same as Fig.\ref{fig:fq_ud_ss} but a factor $p^2 /T^2$ is multiplied.
At $T=0.10$ GeV, it is clear that quarks inside of hadrons have substantially larger occupation probability for high momentum states than the ideal quark gas case. 
This excess in $f^T_{\vq}$ at high momenta is expected to decrease as quarks gradually percolate;
for hadrons greater in size have more chances to accommodate extended, low momentum quark states.
Such extended states in hadrons are partially taken into account through radially excited states with $n_r >1$ 
which contain substantial low momentum states as seen in Fig.\ref{fig:p2_phi^2_ud_ss}.

\subsection{Discussions}
\label{sec:discussions}

\subsubsection{Dissociations of hadrons}
\label{sec:dissociations_hadrons}

We have been interested in how the effective degrees of freedom transform from hadronic to hot and dense phases.
For such transitions to occur smoothly, the properties of quarks in a HRG and a QGP should be close.

We infer that quark states in a HRG tend to have less probability at low momenta but greater probability at higher momenta than in a QGP,
as they are packed into hadrons with the finite size.
Although we still did not include baryons, we have observed such trend at low temperature $T \lesssim 0.15$ GeV in Figs.\ref{fig:fq_ud_ss_T_dep} and \ref{fig:fq_ud_ss_p2}.

Meanwhile, at higher temperatures, $f_{\vq}$ in a HRG begins to exceed the QGP counterpart from low to high momenta. 
We suspect that such overall excess in a HRG is an artifact of extrapolating the ideal gas picture of hadrons.
When hadrons overlap significantly, we expect that quarks percolate, occupying low momentum states.
Based on this picture, we expect the followings to happen:

(i) Hadrons with large radii preferentially merge, and form media filled by color electric fluxes ({\it string condensations} \cite{Polyakov:1978vu}).

(ii) Having colored backgrounds, quarks may be extended in space.
This induces the decay of hadrons at large size into quarks and antiquarks which can get bound to colored backgrounds.
Such percolation enhances the occupation probability at low momenta while reduces the number of high momentum states.

(iii) The decays of highly excited hadrons reduce the hadronic contributions to the thermodynamics,
tempering the drastic growth in thermodynamic quantities, e.g., entropies as in Haged\"orn's gas \cite{Hagedorn:1965st,Broniowski:2004yh}.
Such dissociation also tempers the rapid growth in the quark occupation probabilities.
More precisely, the correct treatment of the decays leads to not only vanishing contributions from decaying hadrons, 
but also results in the {\it negative} contributions to the thermodynamics \cite{Dashen:1969ep,Blaschke:2013zaa,Blaschke:2014zsa,Lo:2017sde,Cleymans:2020fsc}.
Such decay channels open at high energy,
canceling the positive contributions from ordinary resonances at lower energy.
Eventually the thermodynamics is dominated by elementary particles.

We a bit more elaborate the negative contributions just mentioned above. 
For the channel $X$ in two-body problems, the thermodynamic pressure may be expressed using the phase shift $\delta_X$ as
\cite{Dashen:1969ep,Blaschke:2013zaa,Blaschke:2014zsa,Lo:2017sde}
\beq
\hspace{-0.5cm}
P_{\rm th}^X 
=- \int_{\vp} \int_{0}^\infty \frac{\, \rmd \omega \,}{\, \pi \,} T \ln \big( 1-\rme^{- \omega /T} \big) \frac{\, \rmd \delta_X (\omega,\vp) \,}{\, \rmd \omega \,}\,,
\label{eq:P_th_X}
\eeq
where the phase shift obeys the constraint from Levinson's theorem,
\beq
\int_\vp \big[\, \delta_X (\omega =\infty,\vp) - \delta_X (\omega=0, \vp) \,\big]= 0 \,.
\eeq
The physical meaning of this theorem is that interactions cannot change the size of the Hilbert space.
For an ideal meson gas, the phase shift is
\beq
\delta_X^{\rm id} (\omega, \vp) = \sum_n \pi \theta \big(\, \omega - E_n(\vp) \, \big) \,,
\eeq
with which we get the sum of mesonic excitations for Eq.(\ref{eq:P_th_X}).
Actually this expression is inconsistent with the Levinson's theorem, as the $\delta_X$ keeps growing with $\omega$.
The correct phase shift, which carries the information 
that resonances are made of elementary particles,
decreases when the continuum is opened at high energy.
The phase at $\omega=0$ is $\delta_X(\omega=0) = 0$, and must eventually approach zero at $\omega \rightarrow \infty$.
Such domains at high energy show up in thermodynamics at very high temperature, 
yielding the negative contributions which cancel the resonance contributions at lower energy.
For model studies, see, e.g., Refs.\cite{Blaschke:2013zaa,Blaschke:2014zsa} in the context of HRG to QGP transitions.

As we have seen through the Levinson's theorem,
it is in general important to handle double counting of contributions from composite particles and elementary particles.
One of systematic frameworks which automatically handle such double counting is the $\Phi$-derivable approach \cite{Baym:1962sx} 
or the two-particle-irreducible (2PI) formalism \cite{Cornwall:1974vz}.
In particular the zero temperature contributions would be UV divergent unless we properly handle the double counting problem, see, e.g., Ref.\cite{Kojo:2018usw}
in the context of equations of state with quark-hadron continuity;
with proper treatments of double counting, the apparent UV divergences from baryonic contributions and those from quark contributions are assembled to cancel.

\subsubsection{Large $\Nc$, overlap of hadrons, and percolation}
\label{sec:large_nc}

Intuitively the overlap of hadrons is supposed to lead to percolation of quarks, 
but we should be more precise here.
To illustrate the points, here we consider QCD with large number of colors.
In such theories interactions among mesons are $\sim 1/\Nc$ and suppressed \cite{tHooft:1973alw,Witten:1979kh},
while baryons have the masses of $\sim \Nc$ \cite{Hidaka:2008yy}, too massive to participate in the thermodynamics at $T\sim \lqcd$.
The usual conjecture is that the deconfinement happens as the first order transitions,
in the same way as in pure Yang-Mills theories at $\Nc>2$ when gluons saturate the space \cite{Panero:2009tv}.
In pure-Yang Mills at $\Nc=3$, the deconfinement temperature is $T_{\rm YM} \simeq 0.26$ GeV \cite{Boyd:1996bx}, 
considerably larger than the case with quarks, $T\simeq 0.15$ GeV,
where the entropy is $s \sim 2$-3 fm$^{-3}$ indicating the overlap of hadrons with the radii $\sim 1$ fm.

In large $\Nc$,
the spatial overlap of mesons is supposed to be insensitive to $\Nc$, 
but depends only on the size of mesons characterized by $\lqcd$.
Provided that the above argument for meson sizes and $T_{\rm QCD} \rightarrow T_{\rm YM}$ at large $\Nc$,
we have to conclude that the overlap of hadrons are not sufficient to drive the deconfinement or percolation of quarks.
We need a stronger condition.
With meson-meson interactions of  $\sim \Nc^{-1}$,
the transformations of effective degrees of freedom, from hadrons to quarks, happen 
only after the meson abundance becomes large enough to compensate the $1/\Nc$ suppression of interactions.

In order to keep track of quarks from a HGP to a QGP,
it is useful to look at the quark occupation probability in hadrons.
The probability to find a quark with a given color is $\sim 1/\Nc$,
and its magnitude is significantly smaller that  that from percolating quarks where there is no suppression factor of $1/\Nc$. 
This disparity is resolved only when drastic increase in the number of hadrons occurs to compensate the $1/\Nc$ suppression.

Let us estimate the magnitude of meson exchange interactions between mesons,
using the quark-meson vertices $g_{qqM}$ which are known to be $\sim \Nc^{-1/2}$.
At low temperature where the meson abundance is $O(1)$ and hence $f_{\vq} \sim \Nc^{-1}$,
the magnitudes of interactions in a HRG are 
\beq
\hspace{-0.5cm}
\Nc^2 \big( f^{ {\rm low} \,T}_{\vq} \big)^2 g_{qqM}^2 \sim \Nc^2 (\Nc^{-1})^2 ( \Nc^{-1/2} )^2 \sim \Nc^{-1} \,,
\eeq
where we include the combinatorics to pick up quark colors, $\sim \Nc^2$.
Thus the interactions in the thermodynamics are negligible compared to the ideal meson gas contributions of $O(1)$. 

At higher temperature, the interactions become comparable to the ideal gas contributions 
when the meson abundance reaches $O(\Nc)$ due to highly excited states.
(With such large abundance, the ideal gas contributions to the thermodynamic potential also become $O(\Nc)$.)
In the purely mesonic language, the meson three point vertices are $\sim \Nc^{-1/2}$, and the interactions are
\beq
\Nc^2 \big( \Nc^{-1/2} \big)^2 \sim \Nc \,,
\eeq
which is comparable to the ideal meson gas contribution of $\sim \Nc$.
The equivalent descriptions are possible by noting that the meson abundance of $\sim \Nc$ leads to $f^{ {\rm high}\, T}_{\vq} \sim 1$.
Then the magnitude of the interactions among quarks is 
\beq
\hspace{-0.5cm}
\Nc^2 \big( f^{ {\rm high} \,T}_{\vq} \big)^2 g_{qqM}^2 \sim \Nc^2 1^2 ( \Nc^{-1/2} )^2 \sim \Nc \,,
\eeq
which is the magnitude expected from a percolated quark gas.
The corresponding thermodynamic potential is $\sim \Nc$.
Here mesonic and quark gas descriptions should have some overlap in the domain of validity.
This regime holds before the temperature reaches $T_{\rm YM}$ where the thermodynamic potential is $\sim \Nc^2$.
We note that the $f^T_{\vq} $ is bound from above, $f_{\vq}^T \le 1$, so that the quark contribution to the thermodynamics cannot exceed $\sim \Nc$;
diverging behaviors in Haged\"orn's gas are tempered by the quark substructure of hadrons,
at least for resonances made of quarks.

\section{Summary}
\label{sec:summary}

In this work we study a relativistic constituent quark model which is arranged for the studies of the quark-hadron continuity in hot and dense matter.
We analyze mesons including quarks from light to charm quark sectors.
The spectra are reproduced within $\sim $5-10\% accuracy, up to the energy of $\sim 2.5$ GeV.
The impacts of relativistic kinematics, scalar vs vector confinement, and short range correlations are examined.
At conceptual level, considerable differences from the non-relativistic counterpart are found in all these effects.
After obtaining quark wavefunctions, we calculate the occupation probability of quark states in an ideal meson gas.
In order to get insights on the transformation of effective degrees of freedom from a HRG to a QGP, 
we compare the occupation probabilities of quark states in a HRG with those in a QGP.

Although we still have to compute baryonic contributions to complete our HRG descriptions for quarks,
we think that the behaviors of quark occupation probabilities are reasonable in the magnitudes and shapes.
It may be possible to consider a regime where hadrons overlap but the interactions are still tractable in expansion of $1/\Nc$.
Whether one can use such a regime to describe the continuous transformation of effective degrees of freedom is an interesting future subject.

Clearly many aspects in the present paper remain to be improved and extended:

(i) The obvious thing to do is the computations of baryon wavefunctions and the corresponding quark occupation probabilities.
Such computations will complete our descriptions of a HRG.
Furthermore, the quark momentum distributions in a baryon have direct impacts on 
descriptions for cold, dense nuclear matter beyond the nuclear saturation regime \cite{Kojo:2021ugu,Kojo:2021hqh}.

(ii) We think that parameters used for hadron spectra can be more systematically explored,
employing the bayesian or deep learning approaches.
The determination of parameters for inner quark dynamics, e.g., $\alpha_s$, is important by its own.
We should also include more experimental data to get stronger constraints on the model space.

(iii) It is important to calculate the hadron-hadron interaction vertices using the quark wavefunctions. 
The hadronic interactions are related to the underlying hadronic structures which may change in medium.
Such information is relevant in descriptions of baryon rich matters in heavy ion collisions at low and intermediate energies and
in neutron stars.

In addition, the framework in this paper should be also tested in QCD-like theories.
For instance, for QCD in magnetic fields, 
some lattice results and model studies have been available for the structural changes in hadrons \cite{Andreichikov:2016ayj,Hattori:2019ijy,Kojo:2021gvm,Hattori:2015aki,Kojo:2012js,Endrodi:2019whh,Mao:2016fha,Fukushima:2012kc}. 
Also isospin QCD \cite{Son:2000xc,Splittorff:2000mm,He:2005nk,Cohen:2015soa,Brandt:2017oyy,Brandt:2018bwq,Adhikari:2018cea,Adhikari:2019zaj} 
and two-color QCD \cite{Kogut:1999iv,Ratti:2004ra,Brauner:2009gu,Boz:2019enj,Begun:2022bxj,Iida:2019rah,Iida:2020emi} 
have lattice QCD simulations and are important test-beds.
In particular, both mesons and baryons in two-color QCD are calculated within two-body framework, 
so that some predictions can be readily made from the results obtained in the present paper.
The results will be reported elsewhere.

\begin{acknowledgments}

T.K. thanks S. Sasaki and M. Oka for discussions about constituent quark models,
and T. Hatsuda for asking questions about the evolution of the quark occupation probability in finite temperature crossover of QCD.
The work of T.K. was supported by
by the Graduate Program on Physics for the Universe at Tohoku university.


\end{acknowledgments}

\appendix

\section{Coordinate space form factors } \label{app:form_factors}

In Eq.(\ref{eq:k_E_replacement}) we have introduced a form factor for each quark-gluon vertex.
The form factor removes artificial short distance singularities arising from a replacement $\sim \vk/E \rightarrow \vk/\bar{E}$ where the $\vk$-dependence in the denominator is neglected.
The replacement should be valid for $|\vk|\ll \bar{E}$, so we limit the domain of the approximation by multiplying the factors from the vertices 1 and 2,
\beq
F(\vk) =  \rme^{- \vk^2 \big( \Lambda_{v1}^{-2} +  \Lambda_{v2}^{-2}\big) } =  \rme^{- \vk^2 \Lambda_{v}^{-2} } \,,
\eeq
where $\Lambda_{v1,2}$ is $\sim \bar{E}_{1,2}$. 
Meanwhile the large momentum contributions of $|\vk| \gg \bar{E}$ are assumed to be used to renormalize effective model parameters.
Thus such components are simply neglected.

With these form factors, the potentials convoluted with the vertex factors are now treated as
\beq
\!\!\!
 \frac{\, (\rmi k_i \cdots )\,}{\, f( E_1) f(E_2)  \,} U(\vk) 
 ~\rightarrow~ 
\frac{\, (\rmi k_i \cdots ) \,}{\, f( \bar{E}_1) f( \bar{E}_2)  \,} F(\vk)  U(\vk)\,.
 \eeq
Next we take the Fourier transform into the coordinate space expression,
 \beq
&& (\rmi k_i  \cdots ) F(\vk)  U(\vk) ~\rightarrow~ 
 (\partial_i   \cdots ) \int_{ {\bm R} } F(\vecr - {\bm R} )  U( {\bm R} )
\nonumber \\
&&~~~~~~~~~~~~~~ 
=
( \partial_i \cdots )  \int_0^\infty \!\! \rmd R ~ G(r,R) U(R)
 \,,
 \label{eq:deri_GrR_U}
\eeq
with a function regular at small $r$ (we set $\beta = \Lambda_v/2$),
\beq
\!\!\!\!\!
G(r,R) = \frac{\, \beta \,}{\,  \sqrt{\pi} \,}  \frac{\, R \,}{r}
\bigg[\, \rme^{-\beta^2 (r-R)^2 } -  \rme^{- \beta^2 (r+R)^2 } \bigg] \,.
\label{eq:GrR}
\eeq
The derivatives $(\partial_i \cdots)$ hit only $G(r,R)$ so that one can prepare formulae for $(\partial_i \cdots) G(r,R)$ and simply convolute it with any $U(R)$.

The expressions Eqs.(\ref{eq:deri_GrR_U}) and (\ref{eq:GrR}) are already tractable numerically, but we further simplify the expression by extracting the asymptotic behaviors at large $r$ and small $r$, and then by interpolating those analytic expressions.

At small $r$, the first and second terms in $G(r,R)$ tend to largely cancel. For $r \ll \Lambda_v^{-1}$
\beq
 \int_0^\infty \!\! \rmd R ~ G(r,R) U(R) ~ \simeq~ \rme^{ - \beta^2 r^2 } U_{\rm sh}
\eeq
where $U_{\rm sh}$ is the average potential energy in a volume $\beta^{-3}$,
\beq
U_{\rm sh} 
\equiv
  \frac{\, 4 \beta^3 \,}{\,  \sqrt{\pi} \,} \, 
   \int_0^\infty \rmd R\, \rme^{ - \beta^2 R^2 }  R^2 U(R) \,.
\eeq
For $U_{\rm OGE} (r) = - 4\alpha_s/3 r$,
\beq
U_{\rm OGE}^{\rm sh} 
&=& \frac{\, 2 \,}{\, \sqrt{\pi} \,}  \bigg( -\frac{\, 4 \,}{\, 3 \,} \alpha_s \beta  \bigg)
= \frac{\, 2 \,}{\, \sqrt{\pi} \,}  U_{\rm OGE} (\beta^{-1}) 
 \,,
\eeq
and for $U_{\rm conf}^{S,V} (r) = f_{S,V} \times \sigma r$,
\beq
(U_{\rm conf}^{S,V} )^{\rm sh} = \frac{\, 2 \,}{\, \sqrt{\pi} \,} \bigg( f_{S,V} \frac{\, \sigma \,}{\, \beta\,} \bigg) = \frac{\, 2 \,}{\, \sqrt{\pi} \,}  U_{\rm conf}^{S,V} (\beta^{-1}) \,.
\eeq
It turns out that these expressions can be obtained by substituting $\beta^{-1}$ in place of $r$ of the potentials and multiplying a factor $2/\sqrt{\pi}$.

To summarize, our approximation proceeds as follows,
\beq
\!\!\!
 \frac{\, (\rmi k_i \cdots ) \,}{\, f( E_1) f(E_2)  \,} U(\vk) 
 ~\rightarrow~ 
\frac{\, (\rmi k_i \cdots) \,}{\, f( \bar{E}_1) f( \bar{E}_2)  \,} U_{\rm reg} (r) \,,
 \eeq
with $U_{\rm reg} $ interpolating analytic expressions for short and large distance,
\beq
U_{\rm reg} 
= \rme^{- \beta^2 r^2} U_{\rm sh} \,, 
\eeq
as given in Eq.(\ref{eq:def_Uinter}).

\section{ Potentials with derivatives }
\label{app:pot_deri}

We use the following notations for our replacement procedures,
\beq
 (\rmi \vk)_i (\rmi \vk)_j \tilde{U}   ~&\rightarrow&~ 
 \delta_{ij} \bar{U}_2 (r)/3 + Q_{ij}  \bar{U}_{2T} (r)/3  \,,
 \nonumber \\
 ( \rmi \vk \times \vP) \tilde{U} ~&\rightarrow&~ \vL \bar{U}_{1L}  (r)\,, 
 \nonumber \\
  ( \rmi \vk \times \vP)_i   ( \rmi \vk \times \vP)_j \tilde{U} 
 ~&\rightarrow&~ \calP_{ij}  \bar{U}_{2\calP} (r)+ L_i L_j \bar{U}_{2L^2}(r) \,, 
  \nonumber \\
\eeq
where we have defined 
\beq
\calP_{ij} = \delta_{ij} \vP^2 - P_i P_j \,,~~~
Q_{ij} = \delta_{ij} - \frac{\, 3 x_i x_j \,}{\, r^2 \,} \,.
\eeq
The subscript of $\bar{U}$ indicates the powers of $\vk$.
Explicitly, each potential is computed from $U_{\rm reg}$ as
\beq
&&
\bar{U}_2  = {\bm \nabla}^2 U_{\rm reg}\,,~~~~
\bar{U}_{2 T} 
= - \bigg(  \frac{\, \rmd^2  \,}{\, \rmd r^2 \,} - \frac{1}{\, r \,} \frac{\, \rmd \,}{\, \rmd r \,} \bigg)U_{\rm reg} \,,
\nonumber \\
&& ~~~~~~~~~
\bar{U}_{1L} (r) = \bar{U}_{2\calP} (r) 
= \frac{1}{\, r \,} \frac{\, \rmd U_{\rm reg} \,}{\, \rmd r \,} \,, 
\nonumber \\
&& ~~~~~~~
\bar{U}_{2 L^2} (r) 
= \frac{1}{\, r^2 \,} \bigg( \frac{\, \rmd^2  \,}{\, \rmd r^2 \,} -  \frac{1}{\, r \,} \frac{\, \rmd \,}{\, \rmd r \,} \bigg)U_{\rm reg}
\,.
\eeq
It is useful to note 
\beq
{\bm \nabla}^2 U_{\rm reg}  
=
 2\beta^2 (-3+2\beta^2 r^2) \rme^{-\beta^2 r^2} U_{\rm sh}
\,.
\eeq
\beq
 \bigg(  \frac{\, \rmd^2  \,}{\, \rmd r^2 \,} - \frac{1}{\, r \,} \frac{\, \rmd \,}{\, \rmd r \,} \bigg)U_{\rm reg}
= 4 \beta^4 r^2 \rme^{-\beta^2 r^2} \, U_{\rm sh}
 \,.
\eeq
\beq
\frac{\, \rmd U_{\rm reg} \,}{\, \rmd r \,}
= - 2\beta^2 r \rme^{-\beta^2 r^2}  U_{\rm sh} 
 \,.
\eeq
%

\section{Color-flavor-spin factors }
\label{sec:color_flavor_spin_indices}

Here we summarize factors related to the spin and flavor wavefunctions.
For non-flavored mesons, 
we basically assume the ideal mixing which separates $ud$ and $s$-quark sectors, e.g., $\omega \sim u\bar{u} + d\bar{d}$ and $\phi \sim s\bar{s}$.
Only the $\eta$ mesons are treated differently, $\eta \sim u\bar{u} + d \bar{d} - 2s \bar{s}$, as they appear in the low mass region.

First we discuss the factors $b_h^F$.
Here we list only the $b_h^F$'s which are nonzero.
For $\calI=0$ states purely made of $ud$-quarks (e.g., $\omega$),
\beq
 b_{\calI=0,ud}^{u \bar{u} } = b_{\calI=0,ud}^{d \bar{d} } = 1/\sqrt{2} \,.
\eeq
For $\calI=1$ states (e.g., $\pi$, $\rho$),
 \beq
 \hspace{-0.3cm}
 b_{\calI_3=0}^{u \bar{u} } = - b_{\calI_3=0}^{d \bar{d} } = 1/\sqrt{2} \,, ~~~
 b_{\calI_3=1}^{u \bar{d} } = b_{\calI_3=-1}^{d \bar{u} } = 1 \,.
\eeq
For $\calI=1/2$ states (e.g., $K$),
 \beq
 \hspace{-0.3cm}
 b_{\calI_3=1/2}^{u \bar{s} } = b_{\calI_3=1/2}^{s \bar{d} } = 
 b_{\calI_3=-1/2}^{s \bar{u} } =  b_{\calI_3=-1/2}^{d \bar{s} } = 1  \,.
\eeq
For $\calI=0$ states purely made of $s$-quarks (e.g., $\phi$),
 \beq
 \hspace{-0.3cm}
 b_{\calI =0,s}^{s \bar{s} }  = 1  \,.
\eeq
Finally, for $\eta$,
\beq
 \hspace{-0.3cm}
b_{\eta}^{u \bar{u} } = b_{\eta}^{d \bar{d} } = 1/\sqrt{6} \,,
~~~~~ 
b_{\eta}^{d \bar{d} } = -2/\sqrt{6} \,.
\eeq
Next we discuss $N_{\vq SF}^{S_z }$.
Here we pick up $u^R_\up$ for definiteness.
The only nonzero components are
\beq
N_{u^R_\up S (u\bar{u}) }^{S_z } 
= N_{u^R_\up S (u\bar{d}) }^{S_z } 
= N_{u^R_\up S (u\bar{s}) }^{S_z }
= N_{q_\up  S}^{S_z} \,,
\eeq
where
\beq
&& N_{q_\up  S=1}^{S_z=1} = 1\,, ~~~~
N_{q_\up  S=1}^{S_z=-1} = 0\,, ~~~~ 
\nonumber \\
&& N_{q_\up  S=1}^{S_z=0} = N_{q_\up  S=0}^{S_z=0} =1/2\,.
\eeq
Finally we show some examples for terms in our HRG model, Eq.(\ref{eq:Q_in_Jz_sum}).
Setting $\vq=u^R_\up$,
\beq
\hspace{-0.5cm}
 \sum_{J_z} 
 Q_{\rm in}^{h u^R_\up} 
= 
\frac{\, 1 \,}{\, \Nc \,} 
  \sum_{F,S_z}  |b_h^F |^2  \, N_{u^R_\up SF}^{S_z } \,
\frac{\,  | \phi^F_{\vP_h} (p) |^2 \,}{ 4\pi }\,.
\eeq
For $\pi$,
\beq
 Q_{\rm in}^{\pi_0 u^R_\up} 
&=& 
\frac{\, 1 \,}{\, \Nc \,} \frac{1}{\, 2^2 \,}  \, \frac{\,  | \phi^{ud}_{\vP_h} (p) |^2 \,}{ 4\pi } \,,
	\nonumber \\
 Q_{\rm in}^{\pi_+ u^R_\up} 
&=& 
\frac{\, 1 \,}{\, \Nc \,} \frac{1}{\, 2 \,} \,  \frac{\,  | \phi^{ud}_{\vP_h} (p) |^2 \,}{ 4\pi }  \,,
	\nonumber \\
 Q_{\rm in}^{\pi_- u^R_\up} 
&=& 0\,.
\eeq
For $K$, setting $\vq=u^R_\up$,
\beq
&& Q_{\rm in}^{K_+ u^R_\up} 
= 
\frac{\, 1 \,}{\, \Nc \,} \frac{1}{\, 2 \,}  \,\frac{\,  | \phi^{us}_{\vP_h} (p) |^2 \,}{ 4\pi }  \,,
	\nonumber \\
&& Q_{\rm in}^{\bar{K}_0 u^R_\up}  =  Q_{\rm in}^{ K_- u^R_\up} = Q_{\rm in}^{K_0 u^R_\up} 
= 0 \,.
\eeq
or setting $\vq=s^R_\up$,
\beq
 Q_{\rm in}^{K_- s^R_\up} =  Q_{\rm in}^{K_0 u^R_\up} 
&=& 
\frac{\, 1 \,}{\, \Nc \,} \frac{1}{\, 2 \,}  \, \frac{\,  | \phi^{us}_{\vP_h} (p) |^2 \,}{ 4\pi }  \,,
	\nonumber \\
 Q_{\rm in}^{ K_+ s^R_\up} = Q_{\rm in}^{\bar{K}_0 s^R_\up} 
&=& 
0 \,.
\eeq
For $\eta$,
\beq
 Q_{\rm in}^{\eta u^R_\up} 
&=& 
\frac{\, 1 \,}{\, \Nc \,} 
\frac{\,  1  \,}{\, 12 \,} \frac{\,  | \phi^{ud}_{\vP_h} (p) |^2 \,}{ 4\pi }  \, ,
\nonumber \\
 Q_{\rm in}^{\eta s^R_\up} 
&=& 
\frac{\, 1 \,}{\, \Nc \,}  \frac{\, 1 \,}{\, 3 \,} \frac{\,  | \phi^{us}_{\vP_h} (p) |^2 \,}{ 4\pi } 
 \,.
\eeq
For $\rho$,
\beq
\sum_{J_z} Q_{\rm in}^{\rho_0 u^R_\up} 
&=& 
\frac{\, 1 \,}{\, \Nc \,} \frac{3}{\, 2^2 \,}  \, | \phi^{ud }_{\vP_h} (p) |^2 \,,
	\nonumber \\
\sum_{J_z} Q_{\rm in}^{\rho_+ u^R_\up} 
&=& 
\frac{\, 1 \,}{\, \Nc \,} \frac{3}{\, 2 \,} \,  | \phi^{ud }_{\vP_h} (p) |^2 \,,
	\nonumber \\
\sum_{J_z} Q_{\rm in}^{\rho_- u^R_\up} 
&=& 0\,.
\eeq
For $\phi$, we consider $\vq =s^R_\up$,
\beq
\sum_{J_z} Q_{\rm in}^{\phi s^R_\up} 
&=& 
\frac{\, 1 \,}{\, \Nc \,} \frac{3}{\, 2 \,}  \, | \phi^{ss }_{\vP_h} (p) |^2 
\,.
\eeq
As should be already clear, the sum over isospin yields the factor $2\calI+1$.
For $\calI \neq 0$, we can use the formula
\beq
\hspace{-0.9cm}
\sum_{\calI_z, J_z } Q_{\rm in}^{h \vq} 
= \frac{\, (2L+1)(2S+1)(2\calI+1) \,}{\, 4\Nc \,} | \phi^{f }_{\vP_h} (p) |^2 \,.
\eeq
where $f=ud$ or $us$.
For $\calI = 0$ states purely made of $ud$, 
\beq
\hspace{-0.9cm}
\sum_{J_z } Q_{\rm in}^{h \vq} 
= \frac{\, (2L+1)(2S+1) \,}{\, 4\Nc \,} | \phi^{ud }_{\vP_h} (p) |^2 \,,
\eeq
and states purely made of $s$, 
\beq
\hspace{-0.9cm}
\sum_{J_z } Q_{\rm in}^{h \vq} 
= \frac{\, (2L+1)(2S+1) \,}{\, 2\Nc \,} | \phi^{ss }_{\vP_h} (p) |^2 \,.
\eeq
%

%
\section{ Fourier transform } \label{app:fourier}

We discuss how to calculate $\phi_L (k)$ from our coordinate space wavefunctions, including the normalization factor.
Our definition of $\phi_L(k)$ comes from
\beq
\la \vk | \Psi_L^{L_z} \ra = \phi_L(k) Y_L^{L_z}(\hat{\vk}) \,,
\eeq
The normalization condition leads to
\beq
\int_\vk | \Psi_L^{L_z} (\vk) |^2 = 1 ~\rightarrow \int_0^\infty \!\!\! \rmd k \, k^2 |\phi_L(k)|^2 = (2\pi)^3 \,.
\eeq
To relate $\phi_L(k)$ and $\phi_L(r)$, it is convenient to choose $|\vk\ra = | k, {\bm n}_z \ra$ where ${\bm n}_z$ is the quantization axis of $(L,L_z)$.
Then, the angle between $\vk$ and $\vecr$ is the same as the angle appearing in the spherical functions.
We now evaluate\footnote{ 
Our normalizations are $\la \vecr|\vecr'\ra = \delta(\vecr-\vecr')$, $\la \vk |\vk' \ra = (2\pi)^3 \delta(\vk-\vk')$
with which $\la \vecr | \vk \ra = \rme^{-\rmi \vk \cdot \vecr}$ and the closure relations are $1 = \int_{\vecr} |\vecr \ra \la \vecr| = \int_{\vk} |\vk\ra \la \vk|$ with $\int_\vk = \int \rmd \vk/(2\pi)^3$.
The Fourier transformed function as $\phi(\vk) = \int_{\vx} \rme^{\rmi \vk \cdot \vecr} \phi(\vecr)$ and $\phi(\vecr)=\int_{\vk} \rme^{\rmi \vk \cdot \vecr} \phi(\vk)$.
We used $\rme^{\rmi k r \cos \theta} = \sum_l \rmi^l j_l(kr) P_l(\cos \theta)$ 
and $\int \rmd \cos  \theta  P_{L'} (\cos\theta)  P_{L} (\cos\theta) = 2/(2L+1)$.
} 
\beq
&&\la k, {\bm n}_z | \Psi_L^{L_z} \ra
= \int_{\vecr} \la k, {\bm n}_z | \vecr \ra \la \vecr |  \Psi_L^{L_z} \ra
\nonumber \\
&& = 2\pi
\delta_{L_z,0} \sum_{L'} \rmi^{L'} (2L'+1) 
\int \rmd r r^2\, j_{L'} (kr) \phi_L (r)
\nonumber \\
&&~~~~~~~~~
 \times \int \rmd \cos  \theta 
~ P_{L'} (\cos\theta)  P_{L} (\cos\theta)  \sqrt{ \frac{\, 2L+1 \,}{\, 4\pi \,} }
\nonumber \\
&& 
=  \rmi^{L}  \delta_{L_z,0} \sqrt{ 4\pi (2L+1) \,}\, \int \rmd r r^2\, j_{L'} (kr) \phi_L (r) \,.
\label{eq:Qin_sum_Lz}
\eeq
where $j_L$ is the spherical Bessel function and $P_L$ Legendre functions.
Meanwhile
\beq
\la k, {\bm n}_z | \Psi_L^{L_z} \ra 
= \phi_L(k) Y_L^{L_z} ({\bm n}_z) 
=  \phi_L(k)  \delta_0^{L_z}  \sqrt{ \frac{\, 2L+1 \,}{\, 4\pi \,} } \,,
\nonumber \\
\eeq
so that
\beq
\phi_L(k) = 4\pi \int \rmd r r^2\, j_{L} (kr) \phi_L (r) \,.
\eeq

\bibliography{ref}

\end{document}